\documentclass{aa}  
\usepackage{graphicx}
\usepackage{txfonts}
\usepackage{tikz} 
\usepackage[colorlinks=true, citecolor=blue]{hyperref}

\newcommand{\los}[1]{\textit{los}}
\newcommand{\cii}[1]{\textup{[C\textsubscript{II}]}}
\newcommand{\ci}[1]{\textup{[C\textsubscript{I}]}}
\newcommand{\ciidl}[1]{\textup{[C\textsubscript{II}]}\textsubscript{\textup{(DL14)}}}
\newcommand{\ciila}[1]{\textup{[C\textsubscript{II}]}\textsubscript{(L18)}}
\newcommand{\rhoh}[1]{\textit{$\rm \rho$\textsubscript{H2}}}
\newcommand{\zcii}[1]{\textit{z\textsubscript{[CII]}}}
\newcommand{\karcmin}[1]{arcmin\textsuperscript{-1}}
\newcommand{\nuobs}[1]{\textit{$\nu$\textsubscript{obs}}}
\newcommand{\dnu}[1]{\textit{d\textsubscript{$\nu$}}}
\newcommand{\kms}[1]{\textup{km$\cdot$s\textsuperscript{-1}}}
\newcommand{\LciiSRF}[1]{\textit{L\textsubscript{[CII]}}-\textit{SFR}}
\newcommand{\beff}[1]{\textit{b\textsubscript{eff}}}
\newcommand{\bgal}[1]{\textit{b\textsubscript{gal}}}
\newcommand{\bfiteff}[1]{$b^\mathrm{fit}_\mathrm{eff}$}
\newcommand{\bfitgal}[1]{$b^\mathrm{fit}_\mathrm{gal}$}
\newcommand{\pkgal}[1]{\textit{P\textsubscript{gal}}}
\newcommand{\pkj}[1]{\textit{P\textsubscript{J}}}
\newcommand{\klimmpc}[1]{1.35$\times$10$^{-1}$\,Mpc$^{-1}$}
\newcommand{\jup}[1]{$\rm J_{up}$}
\newcommand{\jupref}[1]{$\rm J_{up,ref}$}
\newcommand{\solarmass}[1]{$\rm M_{\odot}$}
\newcommand{\solarlum}[1]{$\rm L_{\odot}$}
\newcommand{\percent}[1]{$\rm \%$}

\usepackage[normalem]{ulem}

\begin{document} 

   \title{Unveiling the evolution of the CO excitation ladder through cross-correlation of CONCERTO-like experiments and galaxy redshift surveys}

   \author{M. Van Cuyck \inst{1} \and
          M. Béthermin \inst{2} \and
          G. Lagache \inst{1} \and
          A. Beelen \inst{1} 
          }

   \institute{Aix Marseille Univ, CNRS, CNES, LAM, Marseille, France
             \and
             Université de Strasbourg, CNRS, Observatoire astronomique de Strasbourg, UMR 7550, 67000 Strasbourg, France  
             }

   \date{Received ; accepted }

 
  \abstract
   {While acting as a foreground for [CII] line intensity mapping (LIM) experiments, rotational CO transitions  serve as valuable indicators of the physical conditions of cold gas in the interstellar media of galaxies at lower redshifts. It is also essential to study these transitions to enhance component separation methods as LIM experiments improve in sensitivity.} 
   {Until now, galaxy-evolution models have been used to predict the total LIM CO signal across all transitions. This study investigates the potential of cross-correlating millimeter-wave LIM data with spectroscopic galaxy surveys to constrain individual CO line contributions. We aim to measure the CO-background spectral-line energy distribution (SLED) and the cosmic molecular gas density \rhoh\,(z) up to z=3.} 
   {We constructed 12 light cones of 9-$\rm deg^2$ from the Simulated Infrared Extragalactic Sky (SIDES) simulation. By analyzing the amplitude ratios of the cross-power spectra between various CO transitions and the galaxy density field, we recovered the CO background SLED. We then combined the derived SLED with the bias-weighted line intensities to estimate \rhoh\,(z). Finally, we assess the feasibility of measuring the CO(4-3) cross-power spectra with a CONCERTO-like experiment.} 
   {Using a realistic depth for a spectroscopic survey, this work accurately recovers the CO background SLED up to \jup\,=6, with uncertainties $\leq$20\%. Then, reconstructing \rhoh, from millimeter-LIM experiments requires introducing an excitation correction relative to CO(1–0). We demonstrate that interloper-induced variance does not prevent precise estimation of \rhoh\,.
   Using the two-star-formation-mode model of SIDES, we show that starbursts contribute significantly to the SLED at \jup\,$\geq$6, but they do not significantly bias the estimation of \rhoh\, from upper 2$\leq$\jup\,$\leq$6 transitions. However, CONCERTO lacks the sensitivity to detect the CO\,x\,galaxy cross-power at relevant scales, even under ideal survey conditions.
}
   {}

   \keywords{Galaxies: evolution – Galaxies: high-redshift - Galaxies: ISM - Cosmology: large-scale structure of the Universe - Cosmology: observations}

\titlerunning{Cross-correlating millimeter-intensity maps with galaxy redshift surveys for CO SLED}
   
   \maketitle
\section{Introduction}

Observations of CO at high redshift with interferometric facilities have advanced our understanding of the cosmic evolution of cool gas and its role in driving the star formation rate density \citep[SFRD,][]{madau14, peroux20, walter20}. Star formation occurs in cold, dense molecular clouds mainly composed of $\rm H_2$, which cannot be directly observed at low temperatures. As the second most abundant molecule and a key coolant of the cold ISM, CO is widely used as a tracer of cold molecular gas \citep{mckee07, carilli13}.

Carbon monoxide emits a ladder of rotational transitions with energies determined by the quantum number \jup,. At high redshift, the ground-state transition CO(1–0) — a key tracer of cold molecular gas — is often not detected, because the signal is very faint and its rest-frame frequency shifts outside the well-covered observational bands.
Consequently, higher-\jup\ lines ($\geq$2) are typically used instead. These must be corrected by an excitation factor before applying the standard CO-to-H$_2$ conversion ($\alpha_{\mathrm{CO}}$). The resulting excitation pattern is the CO spectral-line energy distribution (SLED). Other tracers of the cold ISM include dust continuum \citep[e.g.,][]{scoville17, magnelli20} and atomic lines such as [CI] \citep[e.g.,][]{valentino18, gururajan23}.
In the last ten years, deep CO surveys with NOEMA \citep{boogaard20, boogaard23}, ALMA \citep[ASPECS,][]{decarli20}, and the JVLA \citep[COLDz,][]{riechers19} have measured the molecular gas density, $\rho_{\rm H_2}$, up to $z \sim 5$. These studies reveal an increase by a factor of $\sim7$ in $\rho_{\rm H_2}$ from $z=0$ to $z \sim 1$–3, aligning with the peak of the SFRD \citep{walter20}, and followed by a decline at higher redshift.

Despite uncertainties in the CO-to-H$_2$ conversion factor \citep{bolatto13}, galaxies on the main sequence at $z \sim 1$–3 show higher molecular gas fractions than local ones \citep{tacconi13, tacconi18, tacconi20, saintonge13, genzel15}, suggesting that elevated star formation during cosmic noon is driven by  a higher gas fraction available in galaxies to form stars. 

The shape of the CO SLED reflects molecular gas conditions: low-$J$ lines trace diffuse gas \citep[$\sim10^3\, \rm cm^{-3}$; e.g.,][]{harris10, ivison11, riechers11}, while mid-$J$ lines ($J_{\rm up}=5$–7) trace denser, star-forming regions \citep[$\sim10^5\, \rm cm^{-3}$\, e.g.,][]{daddi15, lu17}. Additional processes such as feedback, shocks, turbulence, active galactic nuclei (AGN) and X-ray heating also affect the SLED. Thus, the SLED provides key insights into the physical conditions of the ISM in galaxies. 
Measured SLEDs can be compared to those of local main-sequence galaxies; starbursts \citep{greve14, rosenberg15, lu17}; high-redshift, dusty star-forming galaxies \citep{danielson11}; and quasars \citep[e.g.,][]{weiss07, bradford09, li20, kade24}. However, the significant variation between individual objects at high redshift makes SLEDs challenging to interpret.
 
Line-intensity mapping (LIM) is a powerful complement to the high-redshift CO observations to obtain a cosmic view of the global CO-emitting properties of galaxies (see the reviews of \citet{kovetz17} and \citet{bernal22}). LIM consists of an accurate mapping of the fluctuations of the CO background intensity produced by the whole galaxy population as a function of sky coordinates and redshift. This observational technique is thus sensitive to CO emitters fainter than the detection limits of galaxy surveys, and thus probes both the clustering of CO emitters and $\rho_{\rm H_2}(z)$, free of any source selection bias. 

Among the various LIM experiments, those operating in the millimeter domain include the South Pole Telescope Summertime Line Intensity Mapper \citep[SPT-SLIM,][,\jup\,$\geq$2]{karkare22} and the CarbON CII line in post-rEionization and ReionizaTiOn epoch \citep[CONCERTO;][\jup\,$\geq$2]{concerto_2020} both targeting \jup,$\geq$2. Other current and forthcoming experiments are conducted in submillimeter wavelengths (primarily targeting the \cii\, line), such as the Fred Young Submillimeter Telescope (FYST)/CCAT-prime with its instrument Prime-Cam \citep[][\jup\,$\geq$2]{aravena23}, the Tomographic Ionized Carbon Intensity Mapping Experiment \citep[TIME;][\jup\,$\geq$2]{crites14, sun21}, and the EXperiment for Cryogenic Large-Aperture Intensity Mapping \citep[EXCLAIM,][\jup\,$\geq$4]{ade20}. These experiments have access to the upper transitions in the CO SLED originating from different redshifts. 

The (sub)millimeter LIM experiments will complement the direct constraints on $\rho_{\rm H_2}(z)$ obtained from radio-LIM experiments that primarily target the CO(1–0) transition, such as COMAP \citep{cleary22}, and will help refine the relation between CO(1–0) emission and the molecular gas mass \citep[e.g.,][]{silva21}. Fluctuations mapped by LIM are typically analyzed using the auto-power spectrum. However, the auto-power spectrum is contaminated by the signal from continuum emission and line interlopers, i.e., other lines from other sources redshifted in the same frequency range. 

On the one hand, one can analyze the total signal formed by the sum of the contribution of the different CO lines from different redshifts with a model. This was carried out by the mmIME experiment \citep{keating20}, which provided the first constraints on total CO shot noise using ALMA and ACT archival data, and derived constraints on the CO(2–1), CO(3–2), and CO(4–3) auto powers at $z$=1.3, $z$=2.5, and $z$=3.6, respectively. \cite{bet22} also compared their empirical CO model to the total CO power constraints of \cite{keating20}, showing agreement with the mmIME measurements.

On the other hand, one can use component-separation methods to obtain information on individual lines directly. In this work, we demonstrate that the cross-correlation between LIM and galaxy surveys can be used to extract the signal from individual CO lines. Cross-correlation between two large-scale structure (LSS) tracers, either between two LIM surveys (e.g., \citet{beane19}, \citet{murmu21}, \citet{schaan21}, \citet{padmanabhan22},  \citet{chung23} and \citet{roy24} for various lines) or with external galaxy surveys (e.g., \citet{wolz17}, \citet{chung19}, and \citet{visbal23}), allows one to isolate the emission originating from the same cosmic volume and to mitigate contamination from interlopers. Indeed, the uncorrelated interloper signal does not contribute on average, but instead acts as an additional source of noise. 

Assuming that continuum contamination from the cosmic infrared background has been removed beforehand (e.g., \citet{yue15} and \citet{vancuyck23})
, we investigated how the CO background SLED and $\rho_{\rm H_2}(z)$ can be constrained from the J$\geq$1 CO transitions up to $z$ = 3.0 using the cross-correlation between LIM data and spectroscopic galaxy surveys.

This paper is organized as follows. In Sect.\,\ref{sec:simu}, we present the fiducial SIDES model on which our study is based. In Sect.\,\ref{sec:cross_pk_mod}, we detail the cross-power spectrum formalism used in this paper. Then, in Sect.\,\ref{sec:ext_crossco}, we present and discuss our results on the reconstruction of the CO background SLED and the cold-gas cosmic density, $\rho_{\rm H_2}$. We assess the feasibility of measuring the cross-power with a CONCERTO-like experiment in Sect.\,\ref{sec:NEI}. Finally, we summarize our conclusions and outline future directions in Sect.\,\ref{sec:ccl}.

\section{SIDES} \label{sec:simu}

Several models of CO line intensity mapping exist in the literature \citep[e.g.,][]{sun21, breysse22, yang22, karoumpis24}.  
Here, we used the simulated infrared dusty extragalactic sky (SIDES) described in \citet{bet17, bet22} and \citet{gkogkou23}, which offers a realistic and observationally motivated framework. SIDES uses a dark-matter halo catalog derived from the 117$\rm deg^2$ light cone of the Uchuu N-body simulation as input \citep{ishiyama21}, achieving the highest mass resolution for a field of this size. SIDES incorporates extensive knowledge of dusty galaxies and the cosmic infrared background (CIB) to reproduce statistical properties across mid-infrared to millimeter wavelengths, including galaxy redshift distributions, number counts, continuum- and line-luminosity functions, CIB fluctuations, SFRD, and the shot-noise level of CO at 3\,mm \citep{keating20}. The SIDES-Uchuu simulation thus provides an optimal compromise between large-field coverage and detailed astrophysical modeling. SIDES adopts the cosmology of the \citet{planck16c} and a \citet{chabrier03} initial mass function (IMF).

\subsection{Galaxy population synthesis}

 The simulated infrared dusty extragalactic sky establishes a relation between stellar mass and halo mass via abundance matching, generating a galaxy light cone complete down to $10^7\,M_\odot$ up to $z = 7$ \citep{gkogkou23}. A scatter of 0.2\,dex is applied to the halo mass–stellar mass relation resulting from the abundance-matching procedure. Similarly, the other observational relations used to derive a galaxy's physical parameters incorporate an appropriate scatter.

The SIDES framework uses a model with two star-formation modes. Galaxies are first classified as star-forming or passive following the observed fraction of star-forming galaxies as a function of mass and redshift from \citet{davidzon17}. Passive galaxies, which are faint in the far-infrared and millimeter wavelengths \citep[e.g.,][]{Whitaker21}, are excluded from further modeling. Star-forming galaxies are assigned to the redshift-evolving main sequence (MS) or to starburst (SB) categories, according to the redshift-dependent fraction described in \cite{bethermin12}, and their SFR is drawn accordingly, with a 0.3\,dex scatter. The SFRs are used to compute infrared luminosities ($L_{\rm IR}$) via the \citet{kennicutt98} relation, which are then used in the line-flux calculation.

\subsection{CO lines}\label{subsec:colines}

The luminosity of the CO(1-0) transition is computed using the $L_{\rm IR} - L_{\rm CO(1-0)}$ relation from \citet{sargent14}, with a scatter of 0.2\,dex. For SB galaxies, which deviate from this relation, an empirical offset of $-0.46$\,dex is applied. SIDES models the SLED of MS galaxies up to J=8 using a combination of clumpy and diffuse gas, with each component adopting a SLED from \citet{bournaud15}. The relative contributions of the clumpy and diffuse gas are set so that the CO(5-4) / CO(2-1) ratios follow the relation of \citet{daddi15}, with the mean radiation field parameter $\langle U \rangle$. In the SIDES framework, $\langle U \rangle$ is used to assign the dust SED and follows a simple redshift evolution that includes a 0.2\,dex scatter: 
\begin{equation}
\log_{10} \left[ \langle U \rangle_{\mathrm{MS}}(z) \right] = \log_{10} \left[ \langle U \rangle_{\mathrm{MS}}(z=0) \right] + \alpha_{\langle U \rangle} \times z \,,
\end{equation}

where $\rm  \langle U \rangle_{\mathrm{MS}}(z=0)$ and $\rm \alpha_{\langle U \rangle}$ are constant. As a result, the scatter in $\langle U \rangle$ causes MS galaxies to exhibit a variety of SLEDs, which generally become more excited with redshift. SB galaxies, in contrast, follow the SLED templates from \citet{birkin_sled}. 

\subsection{Other CO-interloping lines}

The luminosity of the two fine-structure lines \ci\, are derived from empirical relations between luminosity ratios, which were calibrated using the observations from \citet{valentino20}. The calibration of these relations are presented in \cite{bet22}.

First, \citet{valentino20} find that the [CI](1-0) / CO(4-3) is consistent across the different populations of galaxies and redshifts studied. Observations show that the [CI](1-0)/$L_{\rm IR}$ and CO(4-3)/$L_{\rm IR}$ ratios correlate with a dispersion of 0.2dex. This relation is therefore used to derive [CI](1-0). 

The [CI](2-1) / [CI](1-0) ratio traces the excitation temperature. This excitation also influences the CO SLED, making the CO(7-6) / CO(3-4) ratio a good proxy for the excitation temperature in SIDES. These two ratios correlate with a scatter of 0.19\,dex, and they are used to derive the [CI](2-1) line luminosity. 

Finally, [CII] is an important interloper, especially for frequency $\nu>270$\,GHz ($z_\mathrm{[C_{II}]}\leq$6). We adopted the [CII] luminosity from the $L_{\rm [C_{II}]}$-SFR empirical relation calibrated in the local Universe \citep{delooze14} as our fiducial \cii\, luminosity.

\subsection{Spectral cubes} \label{sec:cubes}

From the generated galaxy catalog, mock LIM spectral cubes are produced in sky coordinates for cross-correlation with spectroscopic galaxy surveys. Cross-correlating LIM maps with redshift catalogs isolates emission from the same cosmic volume and mitigates contamination from interloping lines, which contribute additional noise rather than a systematic bias.

The field-to-field variance of the power spectrum impacts millimeter-LIM surveys over small areas, though this effect decreases for larger survey fields \citep{keenan20, gkogkou23}. We therefore extracted 12 light cones of $9\,\rm deg^2$ from the $117\,\rm deg^2$ SIDES-Uchuu simulation to minimize the impact of cosmic variance.

\subsubsection{Sky-coordinate-redshift CO cubes}

For each $9\,\rm deg^2$ field, intensity maps spanning $\delta z = 0.05$ are generated from $z = 0.5$ to $3.0$ for CO transitions with \jup,= 1–8. Table \ref{tab:freqs} lists the observed frequency of each CO line and redshift $z$. The resulting spectral resolutions $\rm \delta\nu$ are given by $\rm \delta\nu = \delta z \times (1+z) \ \nu_{obs}$ and are also provided in Table\,\ref{tab:freqs}, aligning well with the spectral resolution of CONCERTO.

In the intensity maps, intensities from all sources up to $10^7\,M_\odot$ are included because the LIM surveys do not suffer from any selection effect. The emission of interlopers is added to produce interloper-contaminated maps. 

\begin{table*}[h!]
\centering
\caption{Observed frequencies (in GHz) and absolute frequency resolution (in GHz) corresponding to d$z$=0.05 for each CO transition and redshift. } \label{tab:freqs}
\small{\begin{tabular}{p{0.12\linewidth}ccccccc}
\hline
\hline
\noalign{\smallskip}
Line & $\rm \nu_{ref}$ & $\rm \nu_{obs}$ (d$\rm \nu$) & $\rm \nu_{obs}$ (d$\rm \nu$) & $\rm \nu_{obs}$ (d$\rm \nu$) & $\rm \nu_{obs}$ (d$\rm \nu$) & $\rm \nu_{obs}$ (d$\rm \nu$) & $\rm \nu_{obs}$ (d$\rm \nu$) \\
 & & z=0.5 & z=1.0 & z=1.5 & z=2.0 & z=2.5 & z=3.0 \\
\hline
\noalign{\smallskip}
CO(1-0) & 115.3 & 76.8 (2.6) & 57.6 (1.4) & 46.1 (0.9) & 38.4 (0.6) & 32.9 (0.5) & 28.8 (0.4)\\
\noalign{\smallskip}
CO(2-1) & \textbf{230.5} & \textbf{153.7} (5.1) & 115.3 (2.9) & 92.2 (1.8) & 76.8 (1.3) & 65.9 (0.9) & 57.6 (0.7)\\
\noalign{\smallskip}
CO(3-2) & 345.8 & \textbf{230.5} (7.7) & \textbf{172.9} (4.3) & \textbf{138.3} (2.8) & 115.3 (1.9) & 98.8 (1.4) & 86.5 (1.1)\\
\noalign{\smallskip}
CO(4-3) & 461.1 & \textbf{307.4} (10.2) & \textbf{230.5} (5.8) & \textbf{184.4} (3.7) & \textbf{153.7} (2.6) & \textbf{131.7} (1.9) & 115.3 (1.4)\\
\noalign{\smallskip}
CO(5-4) & 576.4 & 384.2 (12.8) & \textbf{288.2} (7.2) & \textbf{230.5} (4.6) & \textbf{192.1} (3.2) & \textbf{164.7} (2.4) & \textbf{144.1} (1.8)\\
\noalign{\smallskip}
CO(6-5) & 691.6 & 461.1 (15.4) & 345.8 (8.6) & \textbf{276.7} (5.5) & \textbf{230.5} (3.8) & \textbf{197.6} (2.8) & \textbf{172.9} (2.2)\\
\noalign{\smallskip}
CO(7-6) & 806.9 & 537.9 (17.9) & 403.4 (10.1) & 322.8 (6.5) & \textbf{269.0} (4.5) & \textbf{230.5} (3.3) & \textbf{201.7} (2.5)\\
\noalign{\smallskip}
CO(8-7) & 922.2 & 614.8 (20.5) & 461.1 (11.5) & 368.9 (7.4) & \textbf{307.4} (5.1) & \textbf{263.5} (3.8) & \textbf{230.5} (2.9)\\
\noalign{\smallskip}
\hline
\end{tabular}}
\label{tab:merged_freq}

\tablefoot{Bold frequencies fall within the band of a CONCERTO-like experiment.}
\end{table*}


\subsubsection{Galaxy number-density-contrast cubes}

To measure the cross-correlation between the CO and galaxies, number density-contrast cubes are generated for each light cone and redshift, spanning the same $\delta$z as the CO cubes.
The depth of the survey used to generate galaxy maps directly impacts the significance of cross-power spectrum detection \citep[e.g.,][]{chung19}. \citet{laigle16} demonstrated that a limiting magnitude of \( K_{s,lim} \leq 24.0 \) ensures 90\% completeness for galaxies with \( M^{\rm L16}_{\rm *,lim} \sim 10^{10} \, \rm M_{\odot} \) at \( z = 2\text{-}3 \) in the COSMOS2015 catalog. Consequently, we include galaxies down to \( 10^{10} \, \rm M_{\odot} \) and up to \( z = 3 \) in our galaxy cubes.

We aim to compute the power spectrum of the galaxy numbe density contrast, $\delta\rho / \rho_0$, where $\rho_0$ is the mean number density. Thus, each galaxy number-density map is normalized by its mean density to produce galaxy number-density contrast maps.

\begin{figure}
    \centering
    \includegraphics[width=0.5\textwidth]{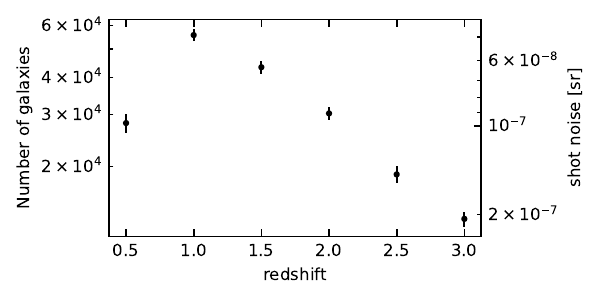}
    \caption{ \textit{Right axis}: Mean number of galaxies per galaxy cube covering 9-deg$^2$ and $\Delta z = 0.25$, as a function of redshift. The error bars come from the dispersion over the twelve independent light cones. \textit{Left axis}: The associated shot noise of galaxy's number density contrast in the 9-deg$^2$ fields.}
    \label{fig:ngal}
\end{figure}

Assuming that the cross-power spectrum remains constant over \(\Delta z = 0.25\), we computed the power spectrum by averaging five spectral channel maps with \(\delta z = 0.05\). Figure\,\ref{fig:ngal} shows the number of galaxies used for cross-correlation, averaged over the 12 light cones in \(\Delta z = 0.25\) as a function of redshift. The right axis of Fig.\,\ref{fig:ngal} shows the corresponding shot noise of the contrast density. 

\begin{figure*}[h!]
        \centering
        \includegraphics[width=1\textwidth]{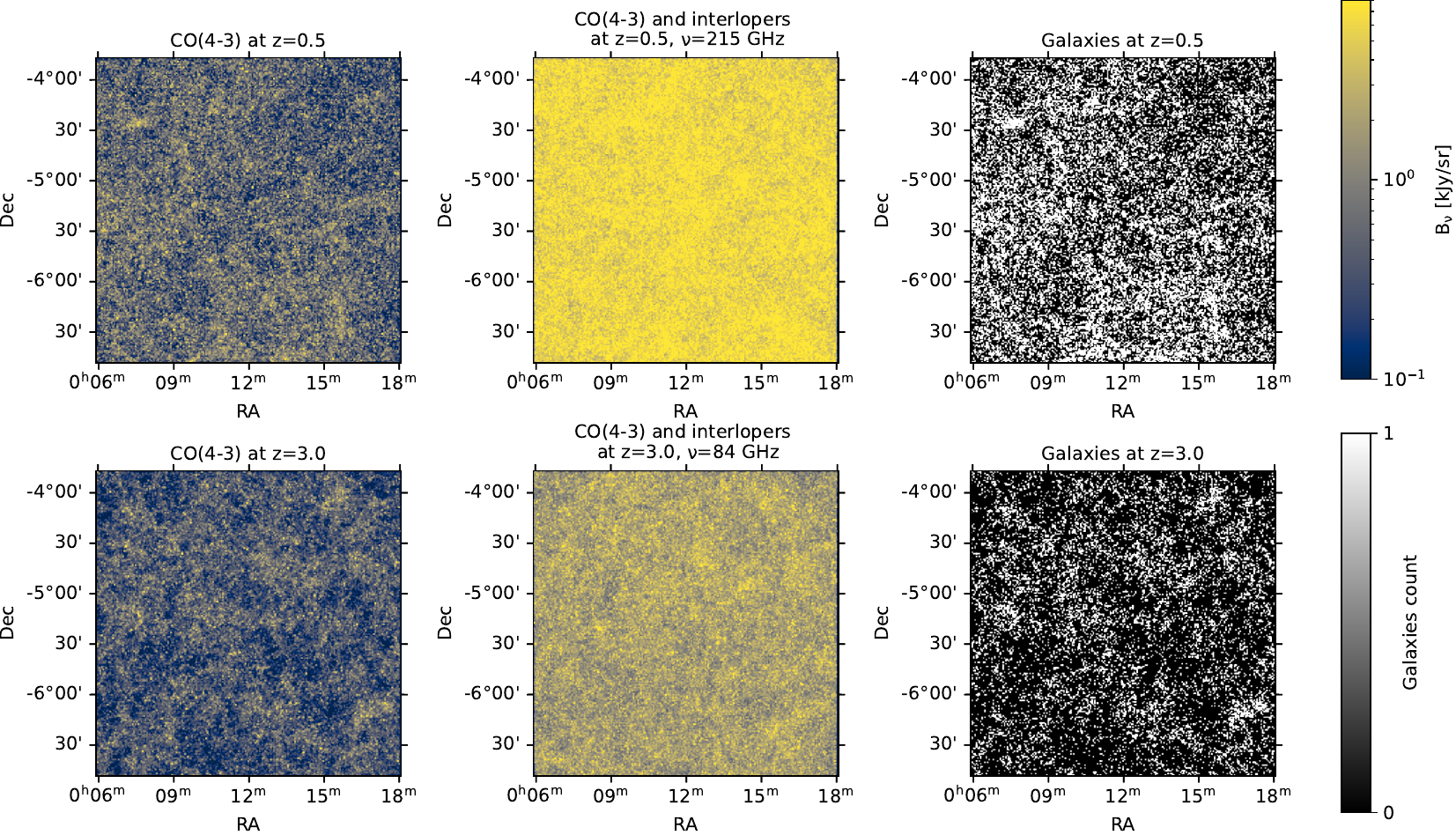}
        \caption{From left to right: Images of SIDES intrinsic CO(4–3) cube integrated along spectral axis over $\Delta z = 0.25$, the same CO(4–3) cube including contamination from line interlopers, and the galaxy cube integrated over the same redshift range. The top row corresponds to \( z = 0.5 \) and the bottom row to  \( z = 3.0 \). Continuum contamination is assumed to have been removed beforehand.}
        \label{fig:comaps}
\end{figure*}

The (cross-) power spectra derived from the mock cubes are called the simulation-based (cross-) power spectra. Figure\,\ref{fig:comaps} shows the CO(4–3) and corresponding galaxy maps, obtained by averaging five maps of $\delta z = 0.05$ along the redshift axis (and thus $\Delta z = 0.25$) for one of $9\,\rm deg^2$ light cones.

\subsection{Relevant scales} \label{sec:revscales}

In this work, we took advantage of the linearity of large scales, where the power spectrum is dominated by the two-halo term of galaxy clustering. To ensure a more robust modeling framework and mitigate the impact of nonlinearities and small-scale systematics, we restricted our modeling and analysis—detailed in the following section—to scales of $k \leq 0.35,\mathrm{Mpc}^{-1}$.
On these scales, the shot noise contribution (strongly dependent on astrophysical details) is subdominant, and instrumental beam smearing has a negligible impact. While the one-halo term could, in principle, contribute, we show later that its impact is negligible on the considered scales ($k\leq$0.35$\rm Mpc^{-1}$).

This choice of scale also constrains the acceptable redshift uncertainty, expressed as dz/(1+z), that an ancillary survey must achieve in order to avoid attenuating the cross-power on the relevant scales.
Following the formalism of \citet[Appendix~B]{chung19}, we find that the typical redshift uncertainty of COSMOS2015, which is $\mathrm{d}z/(1+z) = 3 \times 10^{-2}$, leads to an attenuation of the cross-power spectrum by a factor of $\leq$0.1 at k=0.35$\rm\,Mpc^{-1}$.
By contrast, a redshift uncertainty of $\mathrm{d}z/(1+z) = 1 \times 10^{-3}$, representative of spectroscopic surveys, results in a $\leq$10\% attenuation (a factor of $\geq$0.9) at z$\geq$1, which is smaller than the interloper-induced systematics, as we will see in Sect.\,\ref{sec:ext_crossco}. Increasing to  dz/(1+z)=$3\times10^{-3}$ increases the attenuation to 0.5 at z=1 and to 0.85 at z=3.0. Although current COSMOS2015 and COSMOS2020 do not reach this redshift accuracy, COSMOS is part of the Euclid Wide Survey, which will provide spectroscopic redshifts down to 22.5–23 in AB magnitudes for the bulk of the known COSMOS sources. Therefore, assuming a future spectroscopic survey, we disregarded redshift errors in our galaxy maps for the considered scales: k$\leq$0.35$\rm Mpc^{-1}$.

\section{Power spectrum model}\label{sec:cross_pk_mod}

In this section, we describe the model that we used to interpret the simulation-based angular cross-power spectra measured from the mock cubes described. 

\subsection{Mean background brightness}

The mean background brightness, $B_{\rm \nu}$, at a given observed frequency, $\nu$, produced by point sources is
\begin{eqnarray} \label{eq:intensity_one}
B_{\nu} \, [\rm Jy/sr] = \int S_{\nu}^2 \frac{dN}{dS} d\ln S_{\nu},  
\end{eqnarray}
\noindent where $\rm \frac{dN}{dS}$ are the differential number counts per solid angle, and d$\rm S_{\nu}$ is the observed flux-density interval. The relation between the observed flux density integrated along the line velocity profile $I_{\nu}$ and the spectral luminosity $L_{(1+z)\nu}$ (in L$\rm _\odot$) for a source at a luminosity distance $\rm D_{L}$ in Mpc observed at $\nu$ in GHz is \citep{carilli13}
\begin{equation}\label{eq:S_vs_L}
     I_{\nu} \, [\mathrm{Jy.km.s^{-1}}] =  \frac{1}{1.04 \times 10^{-3} D_{L}^2(z) \nu } \times  L_{(1+z)\nu} \,.
\end{equation}The line flux density, $S_{\nu}$, of the source in a frequency channel measuring $\delta\nu$ in width is then
\begin{equation}\label{eq:snujansky}
    S_{\nu} \,\,\,[\mathrm{Jy}] = \frac{ \nu_{\rm obs}}{c \times \delta{\rm \nu}} \times  I_\nu,
\end{equation}
with the speed of light, $c,$ given in kilometers per second. 

\subsection{Cross-power spectrum}

The clustering of galaxies introduces correlations in the anisotropies. This gives rise to a clustering term in the cross-power spectrum, which encodes the clustered distribution of the line emitters. At large scales, it can be modeled using a simple linear bias model of the following form:

\begin{eqnarray}  \label{eq:cross_clustering}
    P_{\rm J_{up}-gal}^{\rm clust}(k_{\theta},z) = B^{\rm J_{up}}_{\nu}\mathrm(z) b_\mathrm{eff}(z) b_{\rm gal}(z) \times P^\mathrm{2D}_\mathrm{matter}(k_{\theta},z),\,  
\end{eqnarray} 

\noindent where $\rm B_{\nu}$ is given by Eq.\,\ref{eq:intensity_one}. The parameters \beff\, and \bgal\, are the linear clustering biases for the line surface brightness and the galaxies, respectively. As we worked with circular-averaged angular power spectra, $P^\mathrm{2D}_\mathrm{matter}(k_{\theta},z)$ is the 2D, linear, angular power spectrum of matter at the angular wave number $k_{\theta}$.

The 2D angular power spectrum relates approximately to the 3D spherically averaged power spectrum $P^{\rm 3D}(k,z)$ at the comoving wave number $k = \sqrt{(k_{\rm //}^2 + k_{\rm \perp}^2)}$ via \citep{neben17, yue19} 
\begin{equation}\label{eq:pk_matter_2d_3d}
    P^{\rm 2D}(k_{\theta},z) = \frac{1}{\rm D_c^2 \Delta D_c}P^{\rm 3D}\left(k = \frac{2 \pi k_{\theta}}{\rm D_c}, z\right) \,,
\end{equation}
\noindent where $\rm D_{c}$ is the comoving distance in megaparsecs  and $\Delta D_{\rm c}$ is the comoving thickness in the radial direction:
\begin{equation} \label{eq:delta_dc}
    \Delta D_c = \frac{c(1+z) \delta \nu}{H(z) \nu} \,,
\end{equation}
 where $H(z)$ is the Hubble parameter at redshift $z$.
This assumes that $\Delta D_\mathrm{c} << D_\mathrm{c}$ and that $P^{\rm 3D}_{\rm matter}$ does not vary significantly across the frequency channel. The 3D-matter power spectrum, $P_{\rm matter}^{\rm 3D}(z,k),$ was calculated using \textit{CAMB} \citep{camb}. 

The linear clustering bias, \beff\,, quantifies the clustering amplitude ratio between the line surface brightness and the underlying dark-matter distribution on linear scales:
\begin{equation}\label{eq:beff_L}
b_\mathrm{eff}(z) = \frac{\int dM\,n(M,z) b_h(M,z) L(M,z)}{\int dM\,n(M,z) L(M,z)}\, ,
\end{equation}
where n(M,z) is the emitting galaxy's number density and L(M,z) their luminosity given by the SIDES model. The term $\rm b_{\mathrm h}$ is the bias of the halo in which a galaxy lives with respect to the underlying dark-matter distribution.
Since all the galaxies in the same redshift range have similar luminosity distance, \beff\, can be re-expressed in terms of flux density:
\begin{equation}\label{eq:beff_S}
b_\mathrm{eff}(z) = \frac{\int dM\,n(M,z) b_h(M,z) S_\mathrm{\nu}(M,z)}{\int dM\,n(M,z) S_\mathrm{\nu}(M,z)}.
\end{equation}Similarly, the clustering bias for the galaxies, $b_\mathrm{gal}$, is given by
\begin{equation}\label{eq:bgal}
b_\mathrm{gal}(z) = \frac{\int dM\,n_\mathrm{gal}(M,z) b_h(M,z)}{\int dM\,n_\mathrm{gal}(M,z) }
,\end{equation}
where $\rm n_{gal}$(M,z) is the galaxy's number density. 

We ignore the shot noise and one-halo term in our analysis, as it is negligible at the large scales studied here (i.e., below k<0.35$\rm Mpc^{-1}$). The one-halo term primarily dominates on smaller, nonlinear scales, corresponding to the clustering of galaxies within the same dark-matter halo, and its contribution diminishes significantly on larger scales where the two-halo term, representing correlations between halos, becomes dominant. This assumption is supported in the context of CIB analysis, where the one-halo term is reported to be subdominant on these scales \citep[e.g.,][]{viero13}.

\subsection{Auto-power spectra }

The clustering in the auto-power spectrum of the line emitters is

\begin{eqnarray}\label{eq:autocluste}
P_{\rm J_{up}}^{\rm clust}(k_{\theta},z)= B_{\rm \nu}^2(z) b_{\rm eff}^2(z)P^\mathrm{2D}_\mathrm{matter}(k_{\theta},z),\, 
\end{eqnarray}
and the line auto-shot power is given by  
\begin{eqnarray} \label{eq:pshot}
P^\mathrm{shot}(z) = \int S_{\nu}^3 \frac{dN}{dS}d\ln S_{\nu} 
.\end{eqnarray}Similarly, the clustering in the galaxy's auto-power spectrum is modeled as 

\begin{eqnarray} \label{eq:autoclustgal}
P_{\rm gal}^{\rm clust}(k_{\theta},z)=b_{\rm gal}^2(z)P^\mathrm{2D}_\mathrm{matter}(k_{\theta},z)\, ,
\end{eqnarray}
and the galaxy shot noise is given by\begin{eqnarray} \label{eq:gshot}
P^\mathrm{shot}(z) =\frac{1}{\bar{n}_{\rm gal}}
.\end{eqnarray}

\subsection{Clustering biases in SIDES-Uchuu} \label{sec:bfit}

In Eq.\,\ref{eq:cross_clustering}, the mean intensity, $B_{\rm \nu}$, of a line background is degenerated with the clustering bias of CO emitters, \beff\,, and galaxies, \bgal\,. In practice, \bgal\, is measured directly from the galaxy catalog, but \beff\, has to be obtained from a model. Because we work with a simulation, we can derive the effective clustering biases $b^\mathrm{SIDES}_\mathrm{eff}$ and $b^\mathrm{SIDES}_\mathrm{gal}$ specific to the SIDES-Uchuu simulation from the auto-power spectra. 

\begin{figure}
    \centering
    \includegraphics[width=0.5\textwidth]{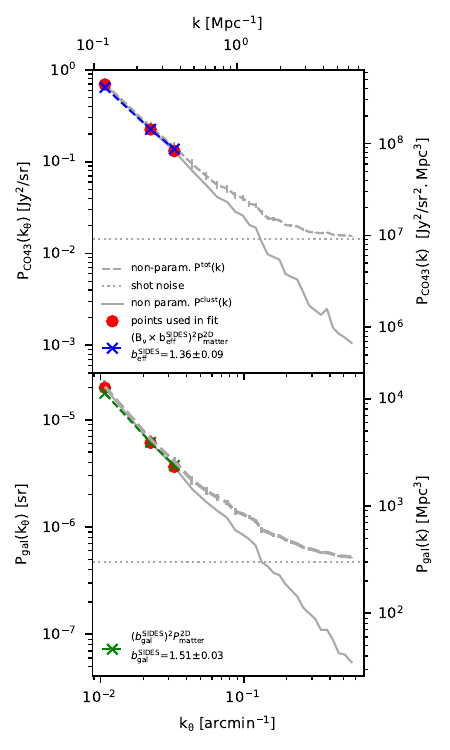}
    \caption{\textit{Upper panel}: CO(4--3) auto-power spectrum at $z=0.5$. The dashed gray line shows the simulation-based auto-power spectrum $\rm P^{tot}(k)$. It is the average over the five auto-power spectra of CO(4--3) maps, each map spanning $\delta z=0.05$. The dotted gray line shows the average shot noise component. Subtracting the shot noise yields the clustering $P^\mathrm{clust}(k)$ shown with the solid gray line. The clustering (Eq.\,\ref{eq:autocluste}) is fit to $P^\mathrm{clust}$ at large scales only, using the red points. From this fit, the effective clustering bias in SIDES, \bfiteff\,, is estimated. Lower panel: Same procedure for galaxies' auto-power spectra and \bfitgal\,, still at $z=0.5$.}
    \label{fig:pkfit}
\end{figure}

To obtain SIDES-specific biases ($b^\mathrm{SIDES}_\mathrm{eff/gal}$) of each CO line at a redshift of $z$ in a subfield, we measured the simulation-based auto-power spectra, $P_\mathrm{J/gal}$. Using the scales below k<0.35$\rm Mpc^{-1}$, we fit the clustering power, $P^\mathrm{clust}_\mathrm{J/gal}$, from Eqs.\,\ref{eq:autocluste} and \ref{eq:autoclustgal}, with $b^\mathrm{SIDES}_{\rm eff/gal}$ as the free parameter to be found. Figure\,\ref{fig:pkfit} shows an example of this procedure at $z=0.5$ for CO(4-3) in one of the 12 subfields. Even though the one-halo term might introduce a slight bias, we demonstrate later that its impact is negligible at the considered scales (k<0.35$\rm Mpc^{-1}$).
 
\section{CO-galaxy cross-correlation \label{sec:ext_crossco}}

In this section, we present the results obtained from the 12 light cones cut in SIDES-Uchuu. For each subfield, we built the spectral cubes at six redshifts from z=0.5 to 3.0, following the method described in Sect.\,\ref{sec:cubes}, as well as the associated galaxy number density-contrast cubes. We cross-correlated each pair consisting of an interloper-contaminated CO intensity cube and a corresponding galaxy density-contrast cube, and measure the simulation-based angular (i.e, in sky coordinates) cross-power spectrum. The results for each redshift and CO transition are averaged over the 12 subfields and presented below. The dispersion among fields is also used to estimate uncertainties. 

In the following, Sect.\,\ref{subsec:sled} explains the reconstruction of the CO background SLED from the measured cross-power spectra and compares to the intrinsic SIDES CO SLED. Sect.\,\ref{subsec:bIs} focuses on the measurement of bias-weighted intensities, and Sect.\,\ref{subsec:beff} focuses on the reconstruction of the effective bias. Finally, Sect.\,\ref{subsec:rhos} discusses the estimation of the cosmic cold-gas density, \rhoh\,($z$), using the results of the two previous sections.

\subsection{Spectral line energy distribution}\label{subsec:sled}

The SLED of the emitter's background is defined as the background intensity ratio between a transition, $\rm J_{up}$, and a reference transition. It can be obtained directly by taking the ratio of two cross-power spectra, to simplify the clustering bias and the matter-power-spectrum factors in Eq.\,\ref{eq:cross_clustering}:

\begin{eqnarray}  \label{eq:sled_from_cross}
    \rm R^{J_{up}}_{J_{up,ref}}(z) &=& \frac{P_{\rm J_{up}-gal}^{\rm clust}(k_{\theta},z) }{P_{\rm J_{up,ref}-gal}^{\rm clust}(k_{\theta},z) } = \frac{B^{\rm J_{up}}_{\nu}\mathrm(z)}{B^{\rm J_{up,ref}}_{\nu}\mathrm(z)} 
    \tikz[baseline=(current bounding box.center)]{
        \node (frac) {\(\frac{b_\mathrm{eff}(z) b_{\rm gal}(z) \times P^\mathrm{2D}_\mathrm{matter}}{b_\mathrm{eff}(z) b_{\rm gal}(z) \times P^\mathrm{2D}_\mathrm{matter}}\)};
        \draw[black, thick] ([xshift=-0.3em]frac.south west) -- ([xshift=0.3em]frac.north east);
    } \,
,\end{eqnarray}
which is only true if $b_\mathrm{eff}$ is the same for all CO lines, and we see that it is indeed the case for SIDES in Sect.\,\ref{subsec:beff}.

\begin{figure}
    \centering
    \includegraphics[width=0.5\textwidth]{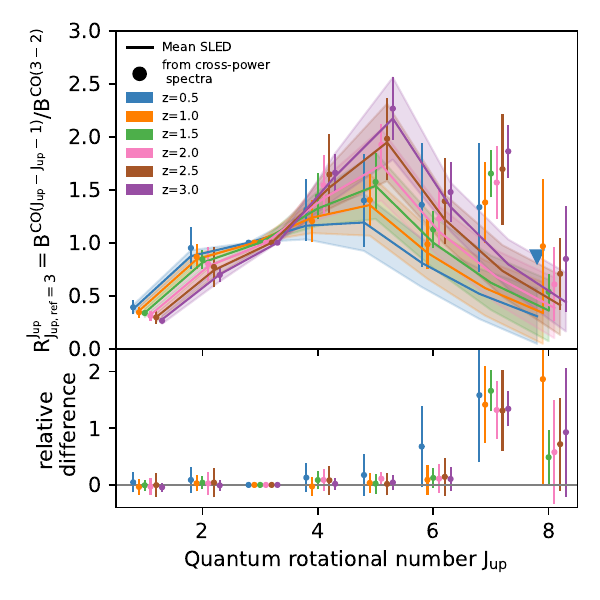}
    \caption{CO SLED and its evolution with redshift. Points have been shifted in the x-axis for clarity. The transition used for normalization is \jup\,=3. Colors encode the redshift, from z=0.5 to 3.0. In the upper panel, solid lines and shaded areas are the mean and $1\sigma$ deviation of the background SLED from the SIDES-Uchuu catalog. Points are obtained from Eq.\,\ref{eq:sled_from_cross} (and thus from the ratio of clustering power spectra) and the associated error bars from the dispersion over the 12 subfields of 9$\rm\,deg^2$. The lower panel shows the relative difference between the mean SLED from the catalog and the measurement from the cross-power spectra.
    \label{fig:sled}}
\end{figure}

Its relevance is twofold. First, the CO background SLED itself, and its evolution with redshift, can provide meaningful information on the physical condition of the cold gas in the interstellar medium of galaxies. Second, deriving the excitation from the SLED is essential to estimating the cold-gas density, \rhoh\,, from CO transitions with \jup\,>1, as the SLED is required to convert the background intensities of these transitions into their equivalent CO(1-0) background intensity. \\
Without access to the \jup\,=1 transition, the millimeter-LIM must first derive the CO SLED and then assume an excitation to connect any level to the \jup\,=1 level. 
In this work, we adopt the CO(3-2) line as the reference, as it allows us to probe up to z=1.5, whereas CO(2-1) is only observable at z=0.5 within the CONCERTO bandwidth. \\
The simulation-based cross-power spectra of all CO lines are measured in each of the 12 simulated light cones at z=0.5, 1.0, 2.0, 2.5, and 3.0. The ratio of the power spectra between a given CO transition and the CO(3-2) line is then taken at each redshift using the scales below k<0.35$\rm Mpc^{-1}$.\\ 
Figure\,\ref{fig:sled} shows with circles the SLED of the CO background obtained from the simulation-based cross-power spectra. We note that the data points have been slightly shifted along the x-axis for clarity. Solid lines and shaded areas represent the mean SLED and its $1\sigma$ dispersion over all galaxies in the SIDES catalog. The color-code corresponds to the six redshift intervals, from $z$=0.5 to 3.0. The lower panel shows the relative difference between the catalog-averaged SLED and the background SLED obtained from the simulation-based cross spectra and Eq.\,\ref{eq:sled_from_cross}.

The latter agrees within the 1-$\sigma$ error bars with the mean SLED for \jup,$\leq$6 at all redshifts. Over all the redshifts, \jup\,=1, 2, 4, 5, 6, and 8 have a relative difference with respect to the mean SLED of 3$\pm$14\%, 4$\pm$17\%, 7$\pm$19\%, 7$\pm$20\%, 22$\pm$32\%, and 75$\pm$85\%, respectively. The relative difference tends to be larger in the z=0.5 bin due to the smaller cosmic volume probed at lower redshift.

Furthermore, the CO(8-7) line is intrinsically faint, leading to non-detection of its clustering signal in two subfields at z=0.5. Hence, we set an upper limit for $\rm R^8_3(z=0.5)$ in Fig.\,\ref{fig:sled}. \\
In addition, we can see that the relative difference evolves with \jup\,. This evolution is linked to the two-star-formation-mode model upon which the SIDES simulation is based, as detailed in Sect.\,\ref{sec:sb}. The peculiar case of CO(7-6), which is overestimated by a factor of $\sim 2$, is discussed in Sect.\,\ref{subsec:J7}.  


\subsubsection{Starbursts contribution to the background intensities} \label{sec:sb}

\begin{figure}
    \centering
    \includegraphics[width=0.5\textwidth]{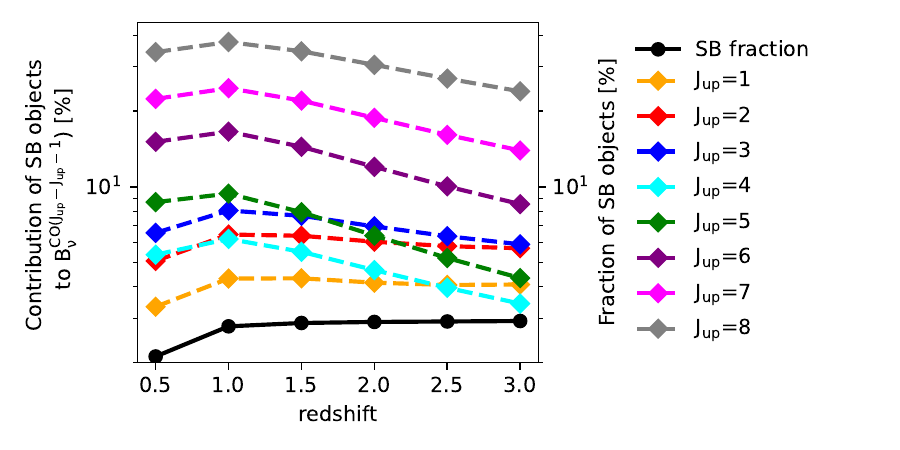}
    \caption{\textit{Left axis and colored dashed curves}: Starburst's contribution to mean brightness of transitions, as function of redshift up to \jup\,=8, inferred from the SIDES-Uchuu catalog. \textit{Right axis and solid black curve}: Starburst fraction in SIDES catalog.} 
    \label{fig:sb_ms}
\end{figure}

The SIDES model is built upon a two-star-formation-mode model from \cite{sargent12}. The catalog contains MS and starburst (SB) galaxies, with an evolving fraction of SB galaxies from $z=0$ to 1 as indicated by the black solid line and the right axis in Fig.\,\ref{fig:sb_ms}. The left axis of Fig.\,\ref{fig:sb_ms} shows the contribution of SB to the mean CO background intensities as a function of redshift. 

We first  examined the evolution of the SB contribution with \jup\,. SB have distinct CO SLED templates compared to MS galaxies, resulting in the SB having a stronger CO emission for \jup\,>5 at fixed stellar and gas mass. Thus, the SB fractional contribution to the background is higher with increasing \jup\,. Although SB galaxies account for at most 3\% of the galaxies in SIDES, their contribution to \jup\,$\leq$5 background intensities is between 3 and 10\,\% over $z$=0.5-3.0. For \jup\,=6 to 8, their contribution rises to $13\pm3$, $20\pm4$, and $31\pm5$\,\%, respectively, on average over $z$=0.5-3.0. 
The low abundance of SB galaxies, combined with their large contribution to the \jup\,=6-8 brightness, leads to the larger variance observed in these CO transition intensities.

Regarding the evolution of the SB contribution with redshift, it reaches its maximum at z=1 for all transitions. Below z=1, the decrease in the SB fraction results in a reduced SB contribution. Above \( z = 1 \), the SB contribution also decreases with increasing redshift, although the SB fraction remains constant and the SB emission template remains fixed with redshift. However, the MS template evolves with \( z \), exhibiting higher excitation. As a result, the SB contribution is lower because the MS galaxies emit more.

We note that only selecting MS galaxies in the catalog would not eliminate the SB contribution in the measured cross-spectra, as the remaining galaxies still trace the large-scale structures where SBs reside. Given the significant contribution of these galaxies to the $B_{\rm \nu}^\mathrm{CO,J_\mathrm{up}>5}$, it is important to give a careful interpretation in terms of global properties, unless these sources can be masked out of the CO maps.

We observe that SB galaxies have no reason to enhance or contribute differently to the effective clustering bias, $\rm b_{eff}$. In SIDES, SB and MS star formation modes are randomly assigned to galaxies that are already clustered. Observationally, \cite{bet14} analyzed COSMOS data and concluded that MS and SB galaxies with the same stellar mass exhibit the same clustering bias and likely reside in halos of similar mass.

\subsubsection{ CO(7--6) contamination by \ci\,(2-1) } \label{subsec:J7}

Figure \ref{fig:sled} shows that the excitations derived for CO(7-6) are largely overestimated at all redshifts. This is caused by the presence of the \ci\,(2-1) line, which originates from the same redshifts and contributes to the cross-power amplitude.  On average, between $z$=0.5 and $z$=3, \ci\, contributes 52\% to the total measured background intensity, with weak evolution across the studied redshift range. Even if the two lines could be resolved, they would still be very close in redshift, practically tracing the same structures. Therefore, cross-correlation could not disentangle the two signals. \\

In summary, the cross-power spectrum method performs well to retrieve the CO background SLED with $\leq$20\%  uncertainty (up to $z$=3.0 and \jup\,=6). The next step is to derive the bias-weighted intensities, which are proportional to \rhoh\,.

\subsection{Effective bias-weighted intensities}\label{subsec:bIs}

From Eq.\,\ref{eq:cross_clustering}, the mean CO-line background intensity weighted by the effective clustering bias of CO emitters can be obtained from the simulation-based measured cross-power spectrum: 
\begin{eqnarray}  \label{eq:bI_from_clustr}
     B^{\rm J_{\rm up}}_\mathrm{\nu}(z) b_\mathrm{eff}(z) =\frac{ P_{\rm J_{\rm up}-gal}^{\rm clust}(k_{\theta},z) } { b^{\rm SIDES}_{\rm gal}(z) \times P^\mathrm{2D}_\mathrm{matter}(k_{\theta},z) } \, ,
\end{eqnarray}
\noindent where $P_{\rm J_{up}-gal}^{\rm clust}(k,z)$ is the clustering component (k<0.35$\rm Mpc^{-1}$) of the simulation-based cross-power spectrum of interloper-contaminated CO maps. 

\begin{figure}
    \centering
    \includegraphics[width=0.5\textwidth]{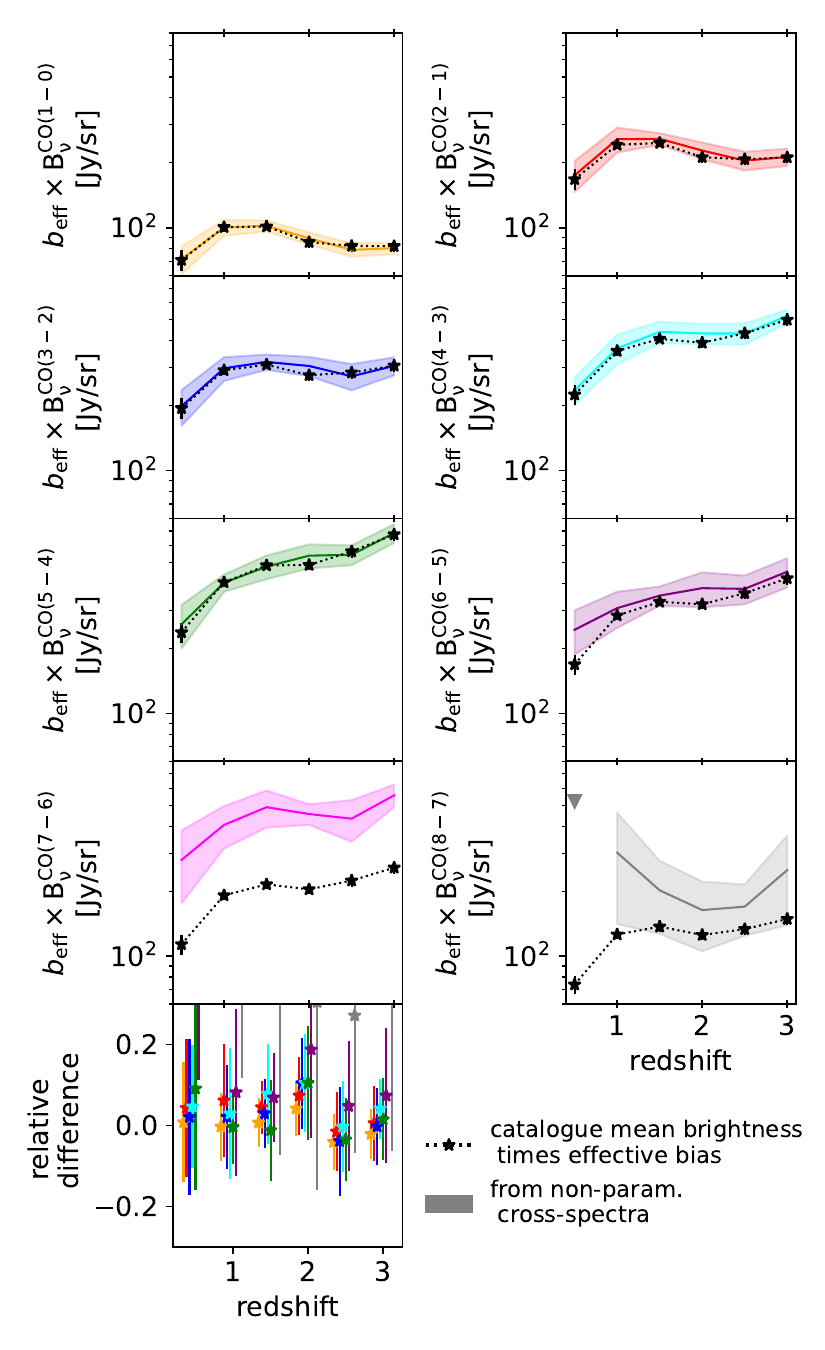}
    \caption{CO bias-weighted intensities as function of redshift. The color-code stands for the transitions' \jup\, from 1 to 8. 
    Solid lines and shaded areas correspond to the mean and 1$\rm \sigma$ dispersion, respectively, on the measured bias-weighted intensities from the 12 $\rm 9\,deg^2$ fields. 
    Black dotted stars show the mean brightness derived analytically times the SIDES effective clustering bias \bfiteff\,.  
    The last panel in the left column shows the relative difference between the simulation-based and analytic bias-weighted intensities (i.e., the relative difference between colored lines and stars).} 
    \label{fig:bI_vs_z}
\end{figure}

Figure\,\ref{fig:bI_vs_z} shows the resulting mean bias-weighted intensities obtained from the cross-power spectra. The color-code corresponds to each CO rotational level (up to \jup\,=8). Solid lines and shaded areas show the mean and $1\sigma$ dispersion from the 12 subfields, respectively.

For a sanity check, the simulation-based bias-weighted intensities are compared to intrinsic values and displayed with stars in Fig.\,\ref{fig:bI_vs_z}. The intrinsic values of $B_\mathrm{\nu}^\mathrm{Jup}(z)$ are obtained from Eq.\,\ref{eq:intensity_one} using the CO flux densities. The bias, $b^{fit}_\mathrm{eff}$, is estimated from the CO line's auto-power spectra, as described in Sect. \ref{sec:bfit}. They constitute "interloper-free" estimates of the bias-weighted intensities in SIDES. The lower panel of the left column in Fig.\,\ref{fig:bI_vs_z} shows the relative difference between the simulation-based and intrinsic values. Apart from CO(7--6), already discussed in Sect.\,\ref{subsec:J7}, simulation-based measured $\rm b_{eff} \times B_{\nu}$ at z$>$0.5 agree with the analytical bias weighted intensities within their 1$\sigma$ uncertainty.

For \jup\,=1 to 5, the 1$\sigma$ uncertainties on $\rm b_{eff} \times B_{\nu}$(z>0.5) from the cross-correlation method are $\rm \pm $10\%. The uncertainties in the z = 0.5 bin are larger, from 10 to 20\%, because this bin probes a smaller cosmic volume. Uncertainties rise to 15-20\percent\, for \jup\,=6. Regarding CO(8-7), we set an upper limit for $\rm b_{eff} \times B^{CO(8-7)}_{eff}$ at z=0.5 in Fig.\,\ref{fig:bI_vs_z} because, as discussed in the previous section, the CO(8-7) cross-spectrum is not detected in two subfields. \\
At z$\geq$1, $\rm b_{eff} \times B^{ CO(8-7)}_{\nu}$ appears to be overestimated by less than $1\sigma$ and shows a variance exceeding 30\%, which is significantly higher than for the other lines.
This overestimation is not significant, and it is due to the intrinsic faintness of the CO(8–7) background combined with significant contamination by the other CO lines and the bright [CII] line, which introduces a substantial level of interloper-induced noise. 
The remaining CO transitions are less affected by [CII] interlopers because of their higher intrinsic brightness. In contrast, the similarly faint CO(1–0) line remains unaffected by [CII] because the latter from $\rm z_{[CII]}\geq 15$ is too faint. In summary, the cross-power-spectrum method effectively recovers the bias-weighted intensities, except for CO(7-6), which is impacted by contamination from [CI]. \\

\subsection{Determination of the effective bias when the background is known} \label{subsec:beff}

We next evaluate our ability to extract the effective bias, \beff\,, from the interloper-contaminated cross-power spectra, given prior knowledge of the background intensities, $B_{\nu}^{\rm J_{up}}$. For this analysis, we assumed that the background level is known from line surveys, although this is typically not the case. Using this information, the goal is to test whether the effective bias can be recovered from the interloper-contaminated cross-power spectra.  

The cross-power spectra are fit in a manner analogous to the auto-power spectra for each CO line, incorporating the known galaxy bias, \bgal\,, and their background intensity, $\rm B_{\nu}$. We then compare the cross-power-derived values of $b_\mathrm{eff}$(z) with those obtained from the auto-power spectra, assessing their consistency and their ability to distinguish among different clustering bias models.

For the model, we used the halo bias model of \citealt{tinker10} (hereafter T10) to compute effective bias:
\begin{equation}\label{eq:modelt10}
    b_\mathrm{eff}(z) = \frac{ \int dM\,n(M,z) b^{\rm T10}_h(M,z) S_\mathrm{\nu}(M,z) }{\int dM\,n(M,z) S_\mathrm{\nu}(M,z)} \, ,
\end{equation}
where $b_h(M,z)$ is the halo clustering bias, and we used its Python implementation from the Colossus package \citep{colossus}. 

We note that the derivation of the analytic clustering bias (Eq.\,\ref{eq:modelt10}) makes use of the flux densities, $S_\mathrm{\nu}$, of the SIDES catalog. Thus, any discrepancy between the auto-power-based and analytic values of $b_{\rm eff}$ reflects the differences between the Uchuu simulation coupled with the abundance-matching procedure of SIDES and the halo-bias model of \cite{tinker10}, which was calibrated on different simulations.

\begin{figure}
    \centering
    \includegraphics[width=\columnwidth]{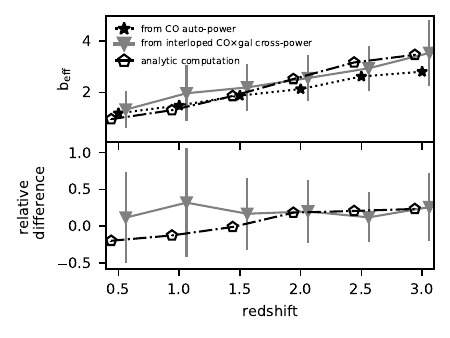}
    \caption{Effective bias of CO-emitters as function of redshift. The points have been slightly shifted in the x-axis for clarity. In the upper panel, stars represent the values fit from the CO auto-power without contamination, as described in Sect.\,\ref{sec:bfit}. Gray triangles are obtained from the fit of interloper-contaminated cross-power spectra between CO LIM and galaxies. Pentagons show the values obtained via the analytic computation using Eq.\ref{eq:modelt10} and the halo bias model of \citet{tinker10}. Values are reported in Table.\,\ref{tab:bias_comparison}. Lower panel shows the relative difference with the auto-power-based values of the bias.} 
    \label{fig:b_vs_z}
\end{figure}

For all three estimates (from auto-power, cross-power, and Eq.\,\ref{eq:modelt10}), the effective bias is the same whatever the CO line (within 3\%), and it is shown as a function of redshift in Fig.\,\ref{fig:b_vs_z}. The consistency of the effective bias across different CO lines, despite the scatter in the SLED of galaxies (see Sect.\,\ref{subsec:colines}), supports the assumption made in Eq.\,\ref{eq:sled_from_cross}. 

At z=0.5, the analytic clustering bias is 18\,\% lower than the SIDES clustering bias. It rises more rapidly and exceeds the SIDES clustering bias at a turnover redshift of approximately z=1.5. At z=3, the analytic clustering bias is 14\,\% higher than the SIDES clustering bias. 

As expected, the error bars on the cross-power-based estimate of $b_{\rm eff}$ are larger due to the interlopers acting as a source of noise, but they agree within their 1-$\rm \sigma$ uncertainty with both the auto-power-based estimate and the analytic computation of the effective bias. Thus, the cross-power method is not inherently biased by interloper contamination; however, their presence limits its ability to constrain the bias model.

\subsection{Cold-gas cosmic density}\label{subsec:rhos}

We now combine our previous results to estimate the cosmic density of cold gas, \rhoh\,(z), and track its evolution with redshift. Converting CO luminosity densities to \rhoh\,(z) implies the use of a global CO-to-H$_2$ mass conversion factor, $\rm \alpha_{\rm CO}$. However, the value of $\rm \alpha_{\rm CO}$ is known to vary with galaxy properties such as metallicity and environment, especially at high redshift. In addition, it tends to be lower in starbursts than in main-sequence galaxies \citep[e.g.,][]{downes98}. As a result, the contributions of different galaxy populations to \rhoh\, depend on their respective abundances, luminosities, and the assumed $\rm \alpha_{\rm CO}$. This introduces an additional layer of degeneracy in the determination of \rhoh\,.

Here, we assign different $\rm \alpha_{\rm CO}$ conversion factors to MS galaxies and starbursts. With this simplified hypothesis, we can test the possibility to recover the gas density from a signal mixing the two populations with different $\rm \alpha_{\rm CO}$. For this exercise, the total molecular gas density we aim to constrain is given by 
\begin{equation} \label{eq:rhoh}
    \rm \rho_{H2}(z) = \rho^{CO(1-0),\,MS}_{L'}(z) \times \alpha^{MS}_{CO} + \rho^{CO(1-0),\,SB}_{L'}(z) \times \alpha^{SB}_{CO} 
,\end{equation}where $\rm \alpha^{MS}_{CO}=4.0$ and $\rm \alpha^{SB}_{CO}=0.8$ in units of $\rm M_{\odot}(K.km.s^{-1}.pc^2)^{-1} $ are the conversion factors for MS galaxies and starbursts, respectively. The terms $\rho^{CO(1-0),\,MS}_{L'}(z)$ and $\rho^{CO(1-0),\,SB}_{L'}(z)$ are the pseudo-luminosity densities of CO(1-0) for these two populations. We computed the molecular gas density and its dispersion from the 12 9-$\rm deg^2$ fields using the SIDES-Uchuu catalog. This computation provides us with a reference to test the accuracy of our method.

To obtain the simulation-based estimates of \rhoh\, from cross-spectra, we combined the simulation-based data obtained in Sects.~\ref{subsec:sled} and \ref{subsec:bIs}, focusing on transitions with \jup\,>1. We began by dividing the bias-weighted CO intensities by the effective clustering biases, as derived from the auto-power spectra in Sect.\,\ref{subsec:bIs}. This step yields the absolute CO background intensities. We then converted these intensities to the equivalent \jupref\,=3 intensities using excitation ratios derived in Sect.\,\ref{subsec:sled}.

Finally, the last step in the \rhoh\, determination for millimeter-wavelength LIM experiments is be to make an assumption about the excitation ratio, $\rm R_{J_{up,ref}-1,}$ between the reference transition (here \jup\,=3) and \jup\,=1 (the latter only being accessible at smaller frequencies with experiments such as COMAP).

We adopted the excitation ratio $\rm R_{3-1}$ from SIDES, which has a value of 2.7 at $z=0.5$ and increases linearly by 33\% between $z=0.5$ and $z=3.0$. This ratio was then used to convert CO(3–2) intensities into their CO(1–0) equivalents.\\

\begin{figure}
    \centering
    \includegraphics[width=0.5\textwidth]{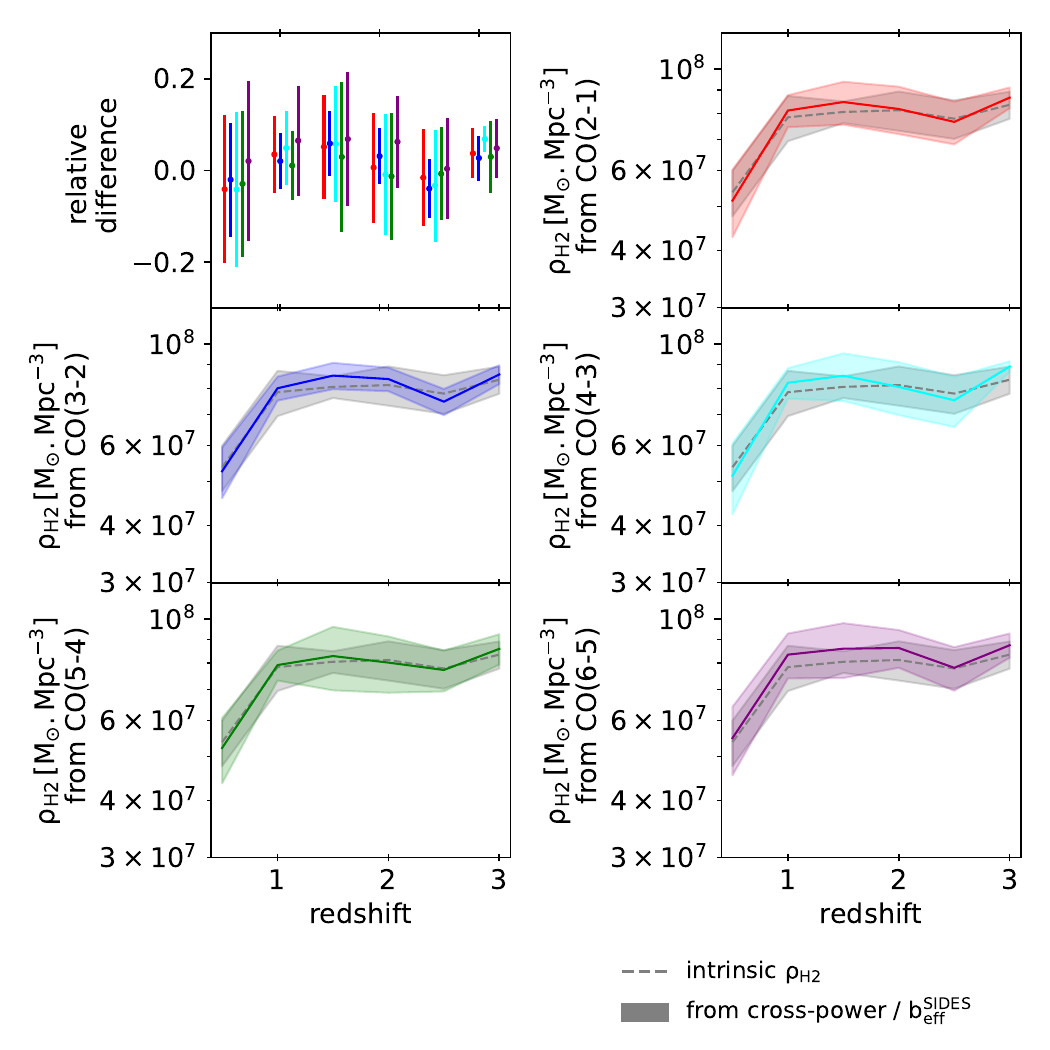}
    \caption{Cold-gas cosmic density, \rhoh\,, as a function of redshift. The gray dashed curve in each panel is the intrinsic \rhoh\, from Eq.\,\ref{eq:rhoh}. Colored solid lines are the simulation-based estimate of \rhoh\, derived from the CO[\jup\,>1] galaxy cross-spectra. The top left panel shows the relative difference from the intrinsic \rhoh\,.} 
    \label{fig:rho}
\end{figure}

The resulting estimates of \rhoh\, are shown in Fig.~\ref{fig:rho} as a function of redshift. For comparison, the intrinsic molecular gas density from Eq.\,\ref{eq:rhoh} is indicated by the gray dashed line in each panel. Transitions with \jup\,=7 and 8 are excluded from the analysis due to contamination by CI and high variance, respectively. The top left panel, where points are offset in the x-axis for clarity, shows the relative difference of \rhoh\, obtained from the simulation-based cross-spectrum estimates and from our reference computed using Eq.\,\ref{eq:rhoh}. 

The mean relative difference between the simulation-based estimates and the reference \rhoh\,(z) values does not exceed 7\% for CO transitions from \jup\,=2 to 6 at all redshifts. The 1-$\rm \sigma$ uncertainties of the relative difference are lower than 20\% for \jup\,=2 to 6, showing only a weak dependence on \jup\,. 

We observe that uncertainties are lower for \jup,=3 (the reference transition), as it does not require estimating excitation ratios relative to higher transitions. The redshift evolution of the CO(3–2) background intensity is similar to that of CO(2–1) and CO(4–3). Thus, changing the reference transition would result in larger uncertainties in the CO(3–2)–based estimates, making them comparable to those derived from CO(2–1) or CO(4–3).

Regarding the influence of SB, the previous section of this paper shows that SB contribution to the background intensity increases with \jup\,. However, this contribution (which could lead to an overestimation of \rhoh\, by boosting the background intensity) is already taken into account in the mean (MS+SB) SLED and so is naturally corrected. 
The contribution of SB to \rhoh\, is less than 1\% and is thus completely negligible compared to that of MS galaxies. However, we emphasize that the SB contribution scales with $\rm \alpha^{SB}_{CO}$, which remains largely uncertain at high redshifts.
Overall, the cross-correlation method with upper CO transitions can constrain the evolution of \rhoh\,(z) to within 20\%. 

Since LIM is sensitive to the integral over the luminosity function, this observational method can bring complementary constraints to line surveys, which are sensitive to the brightest galaxies and rely on extrapolations to account for the faint end of the luminosity function. Data from the (sub)millimeter LIM in particular can complement the CO(1-0) LIM survey. Through cross-correlation, they can be used to constrain the SLED of the CO background across different redshifts. 
Finally, in combination with measurements of the CO(1–0) background, the CO-background SLED offers a valuable opportunity to estimate a "cosmic-averaged" gas-to-mass conversion factor, $\rm \alpha_{CO}$, and its redshift evolution.

\section{Required sensitivity} \label{sec:NEI}
 
Although current LIM experiments typically cover the fields smaller than 9\,deg$^2$ considered above, a survey over 1.5\,deg$^2$ can still, in principle, probe the same linear-scale regime. For instance, a CONCERTO-like experiment over 1.5\,deg$^2$ can measure the power spectrum up to $\rm k_{\theta} > 0.011\,arcmin^{-1}$, corresponding to $k > 0.05\,\mathrm{Mpc}^{-1}$ at $z=1.0$. 

CONCERTO completed its observations at APEX in December 2022, and data reduction of the COSMOS field is currently in progress. An estimate of the on-sky noise-equivalent intensity (NEI) for the spectroscopic mode will become available once this process is finalized, which will in turn enable predictions of the signal-to-noise ratio (S/N) for the CO$\times$galaxy cross-correlation based directly on observational data.

To assess the feasibility of the previous analysis with a CONCERTO-like experiment, we estimated the sensitivity required to detect the CO LIM $\times$ galaxy spectroscopic survey cross-power signal, and compare it with the CONCERTO NEI reported in \citet{concerto_2020} and updated using the instrument characteristics measured on sky \citep{hu24}. CONCERTO has two detector arrays, a first array for low frequencies (LFs) and a second array for high frequencies (HFs). We took the LF array parameters for frequencies below 195\,GHz and the HF parameters at higher frequencies. 

In the above analysis, the cross-power spectrum is measured in two dimensions, first for each individual frequency channel (i.e., redshift slice), and then averaged over five channels. It is well suited because the SIDES simulation offers large fields, dominated by signal. Nevertheless, this 2D estimator is suboptimal with respect to a spherical cross-power that exploits the full set of available modes. 

Thus, we determined the NEI requirements for spherical power spectra and examined whether they could be realistically achieved with a CONCERTO-like experiment. The required NEI to make a significant detection of the cross-power in the two-halo-dominated regime with a given S/N is:
\begin{equation} \label{eq:nei_main}
    \rm NEI = \rm b_{eff} \,B^{J_{up}}_{\nu}\, \frac{\sqrt{\rm t_{survey}\,N_{det}N_{spec}}}{\rm \sqrt{2} \pi\,\,S/N} \,  \left( \int_{k_{\min}}^{k_{\max}} k^2\,P_{\rm matter}^{\rm 3D}(k)\,\mathrm{d}k \right)^{1/2}
\end{equation}
with $k_{\min}=0.1\,\rm Mpc^{-1}$ and $k_{\max}=0.35\,\rm Mpc^{-1}$, which corresponds to the range relevant to the analysis of the previous sections. In this equation, $\rm N_{det}$ and $\rm N_{spec}$ are the number of detectors on the sky for each of the $\rm N_{spec}$ frequency channels of the LIM experiment. The term $\rm t_{survey}$ is the total integration time of the LIM survey. 

The formalism to derive this equation is presented in Appendix\,\ref{app:sensitivity} for a LIM experiment in which the underlying signal is dominated by instrument noise in individual voxel, such as CONCERTO. Here, we focus on the CO(4-3) line, which is covered over a wide redshift range by CONCERTO. The other two bright lines, CO(5-4) and CO(6-5), are discussed in Appendix\,\ref{app:sensitivity}. 

The adopted redshift bins to compute $\rm N_{spec}$ in Eq.\ref{eq:nspec} span $\Delta z$ = 0.5, except at z = 0.5 and z = 2.5, where narrower bins of  $\Delta z$ = 0.23 and 0.32 are adopted, respectively, to take into account that the CONCERTO bandpass does not cover the full redshift bin.
At CONCERTO’s spectral resolution of $\delta\nu=1.5\,\rm GHz$, the radial modes are not sampled up to the full $\rm k_{max}=0.35\,Mpc^{-1}$ range. Therefore we also computed the required NEI taking into account this loss in radial modes. The upper bound of the integration, $\rm k_{max,}$ becomes the larger k mode accessible by an instrument of resolution $\rm \delta \nu=1.5GHz$ at $\rm \nu$ (see Eq.\,\ref{eq:kmax}). It ranges from $\rm k_{max}$=0.26$\,\rm Mpc^{-1}$ at z=0.5 to $\rm k_{max}$=0.13$\,\rm Mpc^{-1}$ at z=2.5.

\begin{figure}
    \centering
    \includegraphics[width=0.5\textwidth]{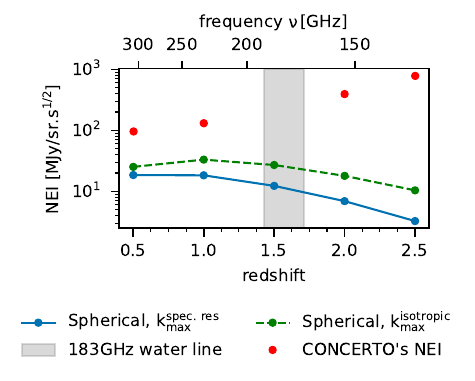}
    \caption{NEI required to detect CO(4-3)\,x-galaxy cross-power with a CONCERTO-like survey at an S/N of 3. The green dashed line corresponds to the NEI needed to detect the spherically averaged cross-power spectrum across the entire $k$ range, [0.1-0.35]$\,\rm Mpc^{-1}$, computed using SIDES. The blue line represents the NEI requirement after accounting for the loss of modes imposed by CONCERTO’s spectral resolution of $\delta \nu=1.5$\,GHz. Estimates for CONCERTO's NEI based on \cite{concerto_2020} and updated with on-sky instrument characteristics from \cite{hu24} are shown in red.}
    \label{fig:NEI}
\end{figure}

Figure\,\ref{fig:NEI} shows the NEI required to detect the CO$\times$galaxy cross-power signal at an S/N of three, together with the estimated CONCERTO NEI. We show two extreme cases of the NEI in this figure. In the first case (dashed green curve), we assumed that all the modes are recovered in the k-range of interest, neglecting the resolution effect. In the second case (solid blue line), we consider that all $k$ larger than the resolution scales are lost, thus neglecting the contribution of transverse and potentially diagonal modes that could be recovered.

Our results show that CONCERTO’s sensitivity is insufficient to constrain the CO(4–3) cross-power, even under optimistic assumptions such as the availability of a spectroscopic galaxy survey (not yet available in the COSMOS field).
Around z$\sim$1.5, the presence of a strong atmospheric water line at 183\,GHz severely limits ground-based observations. 

Our findings draw a consistent picture with the Tomographic Ionized-carbon Mapping Experiment (TIME) forecast by \citet{sun21}, which predicts high S/N values for the CO$\rm \times$galaxy cross-power under the assumption that TIME reaches a NEI of 5$\,\rm MJy \, sr^ {-1} \, s^{^{-1/2}}$ , which is an order of magnitude lower than that required for a CONCERTO-like experiment (Fig.\,\ref{fig:NEI}). Adopting the rest of their survey parameters for CO(4-3) at z=1.0, $\rm N_{feed} \equiv N_{det} =32$, $\rm t_{survey}=1000h$, and taking into account the two-halo term of the clustering in the [0.29-28]$\,\rm Mpc^{-1}$ $k$-range \citep[similarly to that in][]{sun21}, we obtain a high S/N of 90. It is larger than the S/N of 17 that they reported, but the computation in \citet{sun21} takes into account the decorrelation caused by the redshift uncertainty of the galaxy survey (see the discussion in Sect. \ref{sec:revscales}). 

Our results also illustrate how the mode loss impacts the S/N, and consequently, the required NEI. Mode loss due to spectral resolution significantly decreases the goal NEI (solid blue curve), by a factor of $\sim$ 2 at z=0.5 and 10 at z=2.5, compared to when the full $k$-range is exploited (dashed green curve). It highlights that in this noise-dominated regime, fully exploiting all modes (both along and across the line of sight) can significantly boost the S/N of an LIM survey. Thus, to get the most out of cross-correlation, LIM surveys should be designed to sample their scales of interest in the three dimensions well.

\begin{table}[ht] 
\centering
\caption{Adopted CONCERTO parameters to compute NEI in Fig.\,\ref{fig:NEI}.}
\begin{tabular}{lcc}
\hline\hline
\multicolumn{3}{c}{From \citealt{hu24}} \\
\hline
 & LF & HF \\
$N_{\rm det}$ & 1403 & 1183 \\
$\nu_{\rm min}$–$\nu_{\rm max}$ [GHz] & 130–270 & 195–310 \\
$t_{\rm survey}$ [h] & \multicolumn{2}{c}{620} \\

\hline
\hline
 & LF & HF \\
$\rm \nu$ [GHz] & $\rm \nu$<195 & $\rm \nu$$\geq$ 195 \\
$\rm \delta\nu$ [GHz] & \multicolumn{2}{c}{1.5} \\
\hline
\end{tabular}
\label{tab:concerto_params}
\end{table}

\section{Conclusion} \label{sec:ccl}

This study explored the potential of the cross-correlation method between simulated millimeter-line-intensity mapping (millimeter-LIM) data and spectroscopic galaxy surveys to constrain the CO background SLED and the cosmic cold-gas density \rhoh\,(z) from J>1 CO transitions up to z=3.0. The study used SIDES-Uchuu simulations, which offer an optimal compromise between large field coverage and detailed astrophysical modeling.

We used 12 light cones of 9\,deg$^2$ extracted from the SIDES-Uchuu simulation. For each subfield, angular-spectral cubes of CO line intensity were built at six redshifts (from z=0.5 to 3.0 by step of 0.5) as well as the galaxy (density contrast) cubes for a spectroscopic-like survey. We tested the impact of variance induced by interlopers, which can affect the accuracy of the results. Cross-correlation was then performed between each pair of interloper-contaminated CO intensity cube and corresponding galaxy overdensity cube, and the simulation-based angular cross-power spectrum was measured. The results for each redshift and CO transition were then averaged over the 12 subfields.

By analyzing the ratio of two cross-power spectra at different frequencies corresponding to the same redshift for two different CO transitions, the bias and scale-dependent factors can be simplified, enabling constraints on the CO background SLED. This in turn offers valuable insights into the molecular gas density and the physical conditions of the ISM in galaxies. Our main results are as follows:

\begin{itemize}
    
    \item We show that the method is effective in retrieving the CO background SLED and the bias-weighted intensities up to z=3.0 and \jup\,=6. For the SLED, 1$\rm \sigma$ uncertainties are $\leq$20\%.
    The 1$\rm \sigma$ uncertainties on $\rm b_{eff} \times B_{\nu}$ at z$>$0.5 are of $\pm \sim$10\% for \jup\,=1 to 5. This constraint rises to 15-20\% for \jup\,=6 and to >30\% for \jup\,=8.  
    \item CO(7-6) is intrinsically contaminated by [CI], and CO(8-7) appears poorly constrained due to the high dispersion induced by the high interloper contamination.

    \item We find that the effective biases between the different CO lines are the same. Thus, we took the ratio of two CO lines' cross-power spectra corresponding to the same redshift in order to study the global excitation of the CO background. 
    
    \item By combining the results on the CO background SLED and the bias-weighted intensities, we estimated the cosmic density of cold gas \rhoh\,(z) and track its evolution with redshift. The 1$\rm \sigma$ relative uncertainties are lower than 20\% for \jup\,=2 to 6. 

    \item Based on the two-star-formation-mode model of SIDES, we show that the rare starbursts ($\geq$ 3\% up to z$\leq$3) contribute more than 10\% to the background intensity of transitions with \jup\,$\geq$6 (substantially increasing the variance in their cross-power and limiting their interpretation). 
    \end{itemize}

These results demonstrate the effectiveness of the cross-correlation method in isolating line emissions. It also provides important insights into the CO foreground of [CII] above $z \geq 6$, which carries valuable information about star formation in early galaxies \citep[e.g.,][]{liu24}.
This method can bring complementary constraints to the cosmic cold-gas density inferred from CO(1-0), as using the upper CO transitions allows one to constrain a wider range of redshifts from the same LIM dataset.

Our results show that the clustering of the CO cross galaxies power spectrum remains out of reach of a CONCERTO-like experiment in its current configuration. Nonetheless, we also show that leveraging the whole Fourier space, both in the radial and transverse direction to the line of sight, can significantly boost the S/N. It is thus important for a LIM experiment dominated by instrument noise component to sample the scales of interest in the three dimensions sufficiently.

In conclusion, the cross-correlation approach presented here has the potential to improve our understanding of the CO background SLED and the cosmic cold-gas density. It also opens new avenues for using upcoming CO LIM data to probe the interstellar medium and study its role in galaxy evolution.

\bibliographystyle{aa} 
\bibliography{cocrossco} 
\appendix 

\section{Effective bias} \label{app:bias_values}

Table\,\ref{tab:bias_comparison} reports the effective bias values from Fig.\,\ref{fig:b_vs_z}. The first column lists the values obtained analytically using Eq.,\ref{eq:modelt10} together with the halo bias model of \citet{tinker10}. The second column shows the values fitted from the interloper-contaminated CO cross-power, and the third column those fitted from the contamination-free CO auto-power.

\begin{table}[h]
    \small{
    \centering
    \begin{tabular}{c|c|c|c}
        \hline
        Redshift $z$ & Analytic $\rm b_{eff}$ & Cross-power $\rm b_{eff}$ & Auto-power $\rm b_{eff}$\\
        \hline
        \hline
        0.5 & $1.19 \pm 0.07$ & $1.33 \pm 0.73$ & $0.96 \pm 0.01$ \\
        1.0 & $1.49 \pm 0.06$ & $1.97 \pm 1.10$ & $1.31 \pm 0.01$ \\
        1.5 & $1.88 \pm 0.06$ & $2.20 \pm 0.91$ & $1.86 \pm 0.01$ \\
        2.0 & $2.13 \pm 0.06$ & $2.56 \pm 0.92$ & $2.53 \pm 0.01$ \\
        2.5 & $2.63 \pm 0.09$ & $2.94 \pm 0.89$ & $3.17 \pm 0.03$ \\
        3.0 & $2.81 \pm 0.10$ & $3.54 \pm 1.30$ & $3.47 \pm 0.03$ \\
        \hline
    \end{tabular}}
    \caption{Table of effective bias $\rm b_{eff}$ as a function of redshift. }
    \label{tab:bias_comparison}
\end{table}

\section{Sensitivity formalism} \label{app:sensitivity}

In this appendix, we describe the formalism used to compute the sensitivity required to detect the large-scale (linear regime) correlation between galaxy surveys and the CO mapped by a CONCERTO-like experiment. This sensitivity is usually expressed in terms of net equivalent intensity (NEI). In the following equations, the redshift dependence of the variables is neglected for simplicity. This approximation holds if the redshift slice is sufficiently narrow, such that galaxy evolution within it remains negligible. We then apply this formalism to the CO(4-3), CO(5-4) and CO(6-5) lines, which are the brightest lines in the SLED of Fig.\ref{fig:sled}.

\subsection{White-noise power spectrum}

The white-noise power spectrum in a LIM dataset is typically expressed as: 

\begin{equation}\label{eq:noise_auto}
    \rm P^{noise}= V_{voxel} \frac{NEI^2}{t_{int}},
\end{equation}
where $\rm V_{\rm voxel}$ is the comoving volume corresponding to a single voxel (i.e., an element of the spatial–spectral cube), $t_{\rm int}$ is the effective integration time spent on that voxel and the last term is the NEI per detector, which for CONCERTO is frequency dependent. 

Assuming that the LIM experiment observes with an instantaneous field of view $\Omega_{\rm inst}$, in which each detector covers one beam of solid angle $\Omega_{\rm beam}$, the integration time per voxel can be written as:

\begin{equation}
    \rm t_{int} = t_{survey}  \frac{\Omega_{inst}}{\Omega_{survey}} = t_{survey}  \frac{\rm N_{det} \Omega_{beam}}{\Omega_{survey}} = t_{survey} \frac{N_{det}}{N_{beam}},
\end{equation}
where the term $\rm t_{survey}$ is the total integration time on the field and the term $\rm N_{beam} = \Omega_{survey} / \Omega_{beam}$ is the number of independent beams in the LIM field. 

If we approximate that the comoving volume per voxel, $\rm V_{voxel}$, is constant with redshift, it can therefore be expressed as:
\begin{equation}
    \rm V_{voxel} =  \rm V_{survey} \frac{1}{\rm N_{beam} \, N_{spec}},
\end{equation}
where the comoving volume $\rm V_{survey}$ between $(z_{\min}$ and $z_{\max})$ used to measure the cross-power is:
\begin{equation}
    \rm V_{\mathrm{survey}} = \frac{\Omega_{\mathrm{survey}}}{3} \left[ D_c^3(z_{\max}) - D_c^3(z_{\min}) \right],
\end{equation}
We further define $\Delta z$ as $\rm z_{\max} -z_{\min}$. 

The $\rm N_{spec}$ factor corresponds to the number of independent spectral elements within the redshift bin $\Delta z$ and can be computed using:
\begin{equation} \label{eq:nspec}
\rm N_{spec} = \rm (\nu^{z-bin}_{max}-\nu^{z-bin}_{min})/ \delta \nu,
\end{equation}
where $\delta \nu$ is the spectral resolution of the instrument.

Combining the previous equations, we can obtain another expression of the noise auto-power spectrum: 

\begin{equation}\label{eq:noise_auto_2}
    \rm P^{noise} = V_{survey} \frac{NEI^2}{t_{survey}} \frac{1}{N_{det}N_{spec}} 
\end{equation}

\subsection{Signal-to-noise ratio}

Reaching a significance in the detection of the cross-power spectrum, given the survey parameters, implies a required NEI. Here, we want to derive the NEI required to detect the cross-power spectrum between a CO line and a galaxy spectroscopic survey (which is defined in Sect.~\ref{sec:simu}) at a given S/N over the scales of interest. 

We first compute the S/N obtained at a given scale considering all the modes between k and k+$\rm \delta$k:
\begin{equation} \label{eq:dSNR}
\delta \mathrm{S/N}(k) = \frac{\rm P_{J_{\mathrm{up}} \times G}(k)}{ \rm \sigma_{J_{\mathrm{up}}}(k) }
\end{equation}

\noindent where the denominator is the uncertainty in the LIM\,x\,Galaxies cross-power $\rm \sigma_{J_{up}\times G}$ \citep{wolz17}:

\begin{equation} \label{eq:sigmacross}
    \rm \sigma_{J_{\mathrm{up}}}(k) = \frac{\sqrt{ \rm P^2_{J_{up}\times G}(k) + \left( P_{J_{up}}(k) + \rm P^{noise} \right)  P_G(k) }}{ \sqrt{2 \rm \delta N_{modes}(k)}}
\end{equation}

\noindent The term $\rm \delta N_{\rm modes}$ is the number of independent modes between k and k+$\rm \delta$k. 
 
Based on the discussion of Sect.\,\ref{sec:revscales}, we assume that the spectroscopic survey, with dz/(1+z)\,$\sim10^{-3}$, do not introduce an extra attenuation at the relevant scales. We note that using photometric surveys with dz/(1+z)\,$\sim10^{-2}$, available so far in COSMOS, would introduce an non-negligible attenuation of the cross-power amplitudes. 

In the noise-dominated regime characteristic of current LIM experiments, we assume that each voxel is primarily affected by white noise, with only a weak underlying signal. Accordingly, we adopt the approximations $\rm P_{noise} >> P_J(k)$ and $\rm P_{noise}P_G(k) >> P^2_{J-G}(k)$. We verify a posteriori whether these assumptions hold. \\
The resulting goal NEI from these assumptions is the minimal sensitivity that a LIM instrument has to reach to obtain a significant detection. These assumptions simplify Eq.~\ref{eq:dSNR} to:

\begin{equation} \label{eq:SNR}
\delta \mathrm{S/N} =  \frac{\rm P_{J_{\mathrm{up}} \times G}(k)}{\rm \sqrt{\rm P_G(k)}} \frac{\sqrt{\rm  2\,\delta N_{modes}(k)}}{\rm \sqrt{\rm P^{noise}}} 
\end{equation}

At the relevant scales, the two-halo term of the clustering dominates. Thus, using Eq.\,\ref{eq:cross_clustering} and Eq.\,\ref{eq:autoclustgal}, we can simplify the first factor in the previous equation: 
\begin{equation}
    \rm \frac{P_{J_{\mathrm{up}}\times G}(k)}{\sqrt{\rm P_G(k)}} = \rm b_{eff} \,B^{J_{up}}_{\nu} \,\sqrt{\rm P_{\rm matter}^{\rm 3D}(k)},
\end{equation}
where $\rm b_{eff} \,B^{J_{up}}_{\nu}$is the CO line bias-weighted background intensity and $P_{\rm matter}^{\rm 3D}(k)$ is the three-dimensional matter power spectrum, both assumed not to evolve significantly across the redshift bin. We use the value at the center of the redshift bin in our numerical computations. \\

To compute the total S/N, we sum quadratically the $\delta$S/N over all the relevant wavenumbers: 
\begin{equation}
    \mathrm{S/N_{tot}} = \left[ \int \rm \delta S/N^2 \,\rm dk \right]^{1/2},
\end{equation}

The resulting $\mathrm{S/N_{tot}}$ for a given $\rm [k_{min},k_{max}]$ range is then: 
\begin{align}\label{eq:snr_int}
\mathrm{S/N_{tot}} 
&= \sqrt{2}\, \frac{ \rm b_{\mathrm{eff}} \, B^{J_{\mathrm{up}}}_{\nu} }
{ \sqrt{ \rm P^{\mathrm{noise}} } }
\left[ \int_{k_{\min}}^{k_{\max}} \delta N_{\mathrm{modes}}(k)\,
P_{\mathrm{matter}}^{\mathrm{3D}}(k)\, dk \right]^{1/2} \notag \\
&= b_{\mathrm{eff}} \, B^{J_{\mathrm{up}}}_{\nu} \,
   \frac{ \sqrt{ \rm 2\, t_{\mathrm{survey}}\, N_{\mathrm{det}}\, N_{\mathrm{spec}} } }
   { \sqrt{ \rm V_{\mathrm{survey}} } \, \mathrm{NEI} } \notag \\
&\quad \times 
   \left[ \rm \int_{k_{\min}}^{k_{\max}} \delta N_{\mathrm{modes}}(k)\,
   P_{\mathrm{matter}}^{\mathrm{3D}}(k)\, dk \right]^{1/2}
\end{align}
Equation\,\ref{eq:snr_int} indicates that achieving a specific S/N requires attaining a certain level of white noise, which in turn depends on the NEI. In the expression of the S/N, the noise level is balanced by the number of independent modes in the power spectrum measurement.

Given a fix integration time, for all survey sizes that have access to the same given k-range, assuming they have the same galaxy redshift quality, we expect to have the same $\rm S/N_{tot}$. 
However, larger surveys can probe lower $k$ values, thus providing larger S/N if computed using lower $\rm k_{min}$.

\subsection{NEI for spherical cross-power} \label{sec:3d}

In three-dimensional space, the number of independent modes between k and k+$\rm \delta$k is given by:

\begin{equation}\label{eq:nmodes}
    \rm \delta N_{modes} = \frac{1}{2} \frac{V_{survey}}{(2 \pi)^3} \times 4\pi\, k^2\, 
\end{equation}

The 1/2 factor in the expression accounts for the Hermitian symmetry of the Fourier transform of real-valued fields, and considers only the modes in the positive half of the Fourier space. Combining all of the above, the NEI required to significantly detect the spherically averaged cross-power spectrum in the two-halo-dominated regime is: 

\begin{equation} \label{eq:NEIfinal}
    \rm NEI = \rm b_{eff} \,B^{J_{up}}_{\nu}\, \frac{\sqrt{\rm t_{survey}\,N_{det}N_{spec}}}{\rm \sqrt{2} \pi\,S/N} \,  \left( \int_{k_{\min}}^{k_{\max}} k^2\,P_{\rm matter}^{\rm 3D}(k)\,\mathrm{d}k \right)^{1/2}
\end{equation}

The integration is performed on scales of interest of this work, in which the linear clustering of galaxies dominates. The lower integration bound $\rm k_{min}$, is 0.1$\,\rm Mpc^{-1}$, and the upper integration bound $\rm k_{max}$, is 0.35$\,\rm Mpc^{-1}$. 

For consistency, the k-range over which the integral is performed must be well sampled in both the transverse and radial directions to get an isotropic measurement. In the transverse direction, the smallest well-sampled scale (i.e. the largest accessible k) is set by the instrumental beam, and is much smaller than the linear scales used in our analysis. In contrast, in the radial direction, it is set by the spectral resolution. In the radial direction, the smallest scale is the comoving distance between two consecutive frequency channels. Thus, the upper integration bound $\rm k_{max}$ is given by: 
\begin{equation} \label{eq:kmax}
    \rm k_{max} = 2\pi \frac{H \,\nu}{c\,(1+z)\,\delta\nu}\,,
\end{equation}

which, given the CONCERTO-like survey parameter, is smaller than the 0.35$\rm \, Mpc^{-1}$ used in the main text.
Accordingly, we also compute the NEI taking into account this loss of modes by changing the upper integration bound from 0.35$\,\rm Mpc^{-1}$ to the largest accessible mode of Eq.\,\ref{eq:kmax}.

\subsection{CO(5-4) and CO(6-5) NEI}

\begin{figure} 
    \centering
    \includegraphics[width=0.5\textwidth]{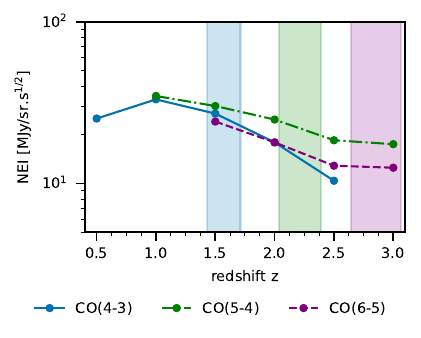}
    \caption{Net equivalent intensity needed to detect the CO LIM cross-galaxies power spectrum at a S/N of 3, assuming an isotropic upper integration bound $\rm k_{max}=0.35\,Mpc^{-1}$. The color code stands for the CO(4-3), CO(5-4) and CO(6-5) lines, in blue, green and purple respectively. Colored areas indicate the redshifts contaminated by the 183\,GHz atmospheric water line for each of the three CO lines. } 
    \label{fig:CO_NEI}
\end{figure}

While the main text focuses on CO(4–3), Fig.\,\ref{fig:CO_NEI} presents the required NEI to detect the cross-power of two other lines: CO(5–4) and CO(6–5), within the CONCERTO frequency band. The redshift bins over which the target NEI is evaluated span $\rm \Delta z = 0.5$, except at $z = 0.5$ and $z = 2.5$ for CO(4–3) and at $z = 1.0$ for CO(5–4). For these cases, the CONCERTO bandpass (130–310\,GHz) does not cover the full redshift bin. Thus, we take narrower bins: $\rm \Delta z = 0.22$ and $0.31$ for CO(4-3) and $\rm \Delta z = 0.37$ for CO(5-4), respectively, in order to match CONCERTO's bandpass.

Overall, the brighter the line, the higher the required NEI, although the order of magnitude remains comparable among the main CO transitions. Consequently, CO(5–4), which corresponds to the peak of the SLED in Fig.\,\ref{fig:sled}, does not require a NEI as low as the two other transitions. 

In particular, at $z = 1.0$, the required NEI is similar for CO(4–3) and CO(5–4), despite the latter being intrinsically brighter. This is due to CO(4–3) falling within the LF array, which has more valid detectors than the HF array hosting CO(5–4) at this redshift (see Table\,\ref{tab:concerto_params}).

\subsection{Validation of Assumptions}

\begin{figure} 
    \centering
    \includegraphics[width=0.5\textwidth]{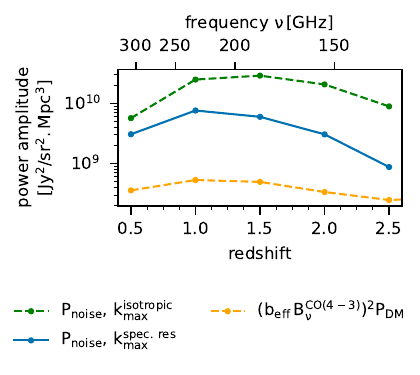}
    \caption{Power amplitudes of the terms entering the expression of $\rm \sigma_{J_{up}-G}(k)$ (Eq\,\ref{eq:sigmacross}). The orange dashed line shows the power amplitude of the cross-term $\rm (b_{eff}\,B_{\nu}^{J_{up}})P^{3D}_{matter}$. The green dashed line shows the white noise power amplitude integrated over the $k$-range 0.1-0.35$\rm \,Mpc^{-1}$. In contrast, the blue solid line shows the white noise power amplitude integrated up to $\rm k_{max}$ defined in Eq.\,\ref{eq:kmax}.     
    Overall, the condition $\rm P_{noise} >> (b_{eff}\,B_{\nu}^{J_{up}})P^{3D}_{matter}$ is verified, but when restricting the integral to the smallest accessible scale in the radial direction, $\rm P_{noise}P_G$ exceeds the cross-term $\rm P_{J_{\mathrm{up}}\times G}$ only by a factor of a few. 
    } 
    \label{fig:verif}
\end{figure}

Using the parameters of Sect.\,\ref{sec:NEI} for a CONCERTO-like survey, we verify that the assumption $\rm P_{noise}P_G(k) >> P^2_{J-G}(k)$ holds. Given the expression of each term, this is equivalent to checking that $\rm P_{noise} >> (b_{eff}\,B^{\nu}_{J_{up}})P^{3D}_{matter}(k)$. Since the noise level depends on the area of the LIM survey, we take $\rm \Omega_{survey} = 1.5\,deg^2$, consistent with the expected size of the CONCERTO field. The term $\rm P^{3D}_{matter}$ is evaluated at k=0.1$\,\rm Mpc^{-1}$. Figure\,\ref{fig:verif} presents the numerical application. When the NEI is computed for the whole $k$-range [0.1,0.35]$\,\rm Mpc^{-1}$, the term $\rm P_{noise}P_G(k)$ is more than one order of magnitude above the cross-term $\rm P^2_{J-G}(k)$. When taking into account the loss in radial modes due to the finite spectral resolution of a CONCERTO-like survey, $\rm P_{noise}P_G(k)$ is a factor of $\geq$9 above the cross term, up to redshift z=2.  At redshift 2.5, $\rm P_{noise}P_G(k)$ still dominates the cross term by a factor of 3, due to the important loss of modes ($\rm k_{max}=0.13\,Mpc^{-1}$ at z=2.5). 

\end{document}